\PassOptionsToPackage{unicode}{hyperref}
\PassOptionsToPackage{hyphens}{url}
\PassOptionsToPackage{dvipsnames,svgnames,x11names}{xcolor}
\documentclass[12pt]{article}

\newcommand{\anon}{1}

\usepackage{amsmath,amssymb,xcolor, amsthm, bm, multirow, url, graphicx, comment, caption, xr, enumitem, booktabs, diagbox, float}
\usepackage{iftex}
\ifPDFTeX
  \usepackage[T1]{fontenc}
  \usepackage[utf8]{inputenc}
  \usepackage{textcomp} % provide euro and other symbols
\else % if luatex or xetex
  \usepackage{unicode-math}
  \defaultfontfeatures{Scale=MatchLowercase}
  \defaultfontfeatures[\rmfamily]{Ligatures=TeX,Scale=1}
\fi
\usepackage{lmodern}
\ifPDFTeX\else  
\fi
\IfFileExists{upquote.sty}{\usepackage{upquote}}{}
\IfFileExists{microtype.sty}{% use microtype if available
  \usepackage[]{microtype}
  \UseMicrotypeSet[protrusion]{basicmath} % disable protrusion for tt fonts
}{}
\makeatletter
\@ifundefined{KOMAClassName}{% if non-KOMA class
  \IfFileExists{parskip.sty}{%
    \usepackage{parskip}
  }{% else
    \setlength{\parindent}{0pt}
    \setlength{\parskip}{6pt plus 2pt minus 1pt}}
}{% if KOMA class
  \KOMAoptions{parskip=half}}
\makeatother
\usepackage{xcolor}
\makeatletter
\ifx\paragraph\undefined\else
  \let\oldparagraph\paragraph
  \renewcommand{\paragraph}{
    \@ifstar
      \xxxParagraphStar
      \xxxParagraphNoStar
  }
  \newcommand{\xxxParagraphStar}[1]{\oldparagraph*{#1}\mbox{}}
  \newcommand{\xxxParagraphNoStar}[1]{\oldparagraph{#1}\mbox{}}
\fi
\ifx\subparagraph\undefined\else
  \let\oldsubparagraph\subparagraph
  \renewcommand{\subparagraph}{
    \@ifstar
      \xxxSubParagraphStar
      \xxxSubParagraphNoStar
  }
  \newcommand{\xxxSubParagraphStar}[1]{\oldsubparagraph*{#1}\mbox{}}
  \newcommand{\xxxSubParagraphNoStar}[1]{\oldsubparagraph{#1}\mbox{}}
\fi
\makeatother

\usepackage{longtable,booktabs,array}
\usepackage{calc} % for calculating minipage widths
\usepackage{etoolbox}
\makeatletter
\patchcmd\longtable{\par}{\if@noskipsec\mbox{}\fi\par}{}{}
\makeatother
\IfFileExists{footnotehyper.sty}{\usepackage{footnotehyper}}{\usepackage{footnote}}
\makesavenoteenv{longtable}
\usepackage{graphicx}
\makeatletter
\def\maxwidth{\ifdim\Gin@nat@width>\linewidth\linewidth\else\Gin@nat@width\fi}
\def\maxheight{\ifdim\Gin@nat@height>\textheight\textheight\else\Gin@nat@height\fi}
\makeatother
\setkeys{Gin}{width=\maxwidth,height=\maxheight,keepaspectratio}
\makeatletter
\def\fps@figure{htbp}
\makeatother

\makeatletter
\@ifpackageloaded{caption}{}{\usepackage{caption}}
\AtBeginDocument{%
\ifdefined\contentsname
  \renewcommand*\contentsname{Table of contents}
\else
  \newcommand\contentsname{Table of contents}
\fi
\ifdefined\listfigurename
  \renewcommand*\listfigurename{List of Figures}
\else
  \newcommand\listfigurename{List of Figures}
\fi
\ifdefined\listtablename
  \renewcommand*\listtablename{List of Tables}
\else
  \newcommand\listtablename{List of Tables}
\fi
\ifdefined\figurename
  \renewcommand*\figurename{Figure}
\else
  \newcommand\figurename{Figure}
\fi
\ifdefined\tablename
  \renewcommand*\tablename{Table}
\else
  \newcommand\tablename{Table}
\fi
}
\@ifpackageloaded{float}{}{\usepackage{float}}
\floatstyle{ruled}
\@ifundefined{c@chapter}{\newfloat{codelisting}{h}{lop}}{\newfloat{codelisting}{h}{lop}[chapter]}
\floatname{codelisting}{Listing}

\makeatother
\makeatletter
\@ifpackageloaded{caption}{}{\usepackage{caption}}
\@ifpackageloaded{subcaption}{}{\usepackage{subcaption}}
\makeatother

\ifLuaTeX
  \usepackage{selnolig}  % disable illegal ligatures
\fi
\usepackage[]{natbib}
\usepackage{bookmark}

\IfFileExists{xurl.sty}{\usepackage{xurl}}{} % add URL line breaks if available
\hypersetup{
  pdftitle={Title},
  pdfauthor={Author 1; Author 2},
  pdfkeywords={3 to 6 keywords, that do not appear in the title},
  colorlinks=true,
  linkcolor={blue},
  filecolor={Maroon},
  citecolor={Blue},
  urlcolor={Blue},
  pdfcreator={LaTeX via pandoc}}

\newcommand{\rexp}{\mathrm{Exp}}
\newcommand{\rlog}{\mathrm{Log}}
\newcommand{\D}{\mathrm{d}}
\newtheorem{theorem}{Theorem}
\newtheorem{prop}{Proposition}%
\newtheorem{example}{Example}%
\newtheorem{remark}{Remark}%

\begin{document}

\def\spacingset#1{\renewcommand{\baselinestretch}%
{#1}\small\normalsize} \spacingset{1}

%%%%%%%%%%%%%%%%%%%%%%%%%%%%%%%%%%%%%%%%%%%%%%%%%%%%%%%%%%%%%%%%%%%%%%%%%%%%%%

\if1\anon
{
  \title{\bf A Riemannian Factor Model\\ for Manifold-Valued Time Series%\thanks{
    %Rong Chen's research was supported in part by National Science Foundation grants DMS-2319260 and DMS-2413858. Yaqing Chen's research was supported in part by National Science Foundation grant DMS-2311035.}
    }
  \author{Shuo-Chieh Huang\\
    Department of Statistics, Rutgers University\\
    Rong Chen \\
    Department of Statistics, Rutgers University \\
    and \\
    Yaqing Chen \\
    Department of Statistics, Rutgers University}
    \date{}
  \maketitle
} \fi

\if0\anon
{
  \begin{center}
    \LARGE\bf A Riemannian Factor Model\\ for Manifold-Valued Time Series
\end{center}
  \medskip
} \fi

\bigskip
\begin{abstract}
We propose a Riemannian factor model (RFM), a novel framework for analyzing potentially high-dimensional time series data observed on Riemannian manifolds. Such time series are encountered in various applications, including economics, finance, medical imaging, and genomics and microbiome research. The proposed model is geometry-aware and accounts for the inherent nonlinearity in the data. In a high-dimensional asymptotic regime, where the manifold dimension is allowed to diverge with the sample size $n$, we establish convergence rates for the estimated loading space. In particular, under short-memory and strong factor conditions, we obtain a dimension-free $n^{-1/2}$ rate, which matches the convergence rate of the high-dimensional linear factor model. Finite-sample performance of the proposed RFM is demonstrated with simulated time series on the Bures--Wasserstein manifolds and products of spheres, as well as an application to monthly realized covariances of selected U.S. stock returns---modeled as time series in the Bures--Wasserstein manifold, where the RFM provides demonstrably interpretable factors and yields competitive predictive performance.
\end{abstract}

\noindent%
{\it Keywords:} Bures--Wasserstein metric, compositional time series, covariance prediction, dimension reduction, non-Euclidean time series
\if1\anon{
\vfill
}\fi

% \newpage
\spacingset{1.8} % DON'T change the spacing!

\section{Introduction}
In recent years, time series that intrinsically live on a nonlinear manifold have emerged as an important type of data in many scientific fields. For example, compositional time series can be treated as sphere-valued data after the square root transformation \citep{WANG2007, dai2018, zhu:24}. As another example, distribution-valued data, such as the distributions of age-at-death, housing price \citep{Chen2023}, stock returns \citep{Zhang2022} and the spatial distribution of taxi demand \citep{Dai2022}, are also naturally nonlinear. Similarly, observations of symmetric positive-definite (SPD) matrices, common in medical imaging (e.g., diffusion tensor imaging), can be analyzed as data on some nonlinear manifold \citep{Moakher2005, Pennec2006, Arsigny2007, Harandi2018}. SPD-valued data are also ubiquitous in finance, where time-varying covariance matrices play a key role \citep{Kasper2023}.

%{[For the following paragraph, I suggest we start with manifold-valued data, of which the analysis is often faced with the difficulty of high dimensionality. Hence dimension reduction is needed. Then state the existing work for independent Riemannian data (see \url{https://arxiv.org/pdf/2602.05936} for a review) and Riemannian functional data. Then point the gap---no work on Riemannian time series. Then start a new paragraph for factor models and what follows.]} 

In addition to the intrinsic nonlinearity, the analysis of manifold-valued time series is often faced with high dimensionality. 
Indeed, dimension reduction of non-Euclidean data has been a topic of great interest and was studied under a few different settings. 
For instance, \citet{Fletcher2004} proposed the principal geodesic analysis (PGA), which performs linear PCA on the tangent space of the Fr\'{e}chet mean---a generalization of the mean to metric space-valued random elements \citep{frechet1948}.
\citet{Huckemann2006} considered alternative principal geodesics that typically do not pass the Fr\'{e}chet mean.
For specialized nonlinear spaces, \citet{Jung2012} proposed the principal nested spheres (PNS) for data on hyperspheres and \citet{Huckemann2010} considered quotient spaces from isometric Lie group actions.
See also \citet{Huckemann2021} for a discussion on the difficulty of extending PNS to other manifolds.
\citet{dai2018} and \citet{Lin2019} considered dimension reduction for functional observations on a Riemannian manifold. 
Nevertheless, the aforementioned works either focus on independent samples of functions or lack statistical guarantees. 
As a result, the rigorous development of a dimension reduction framework for Riemannian time series remains an important gap in the current literature.

Factor models \citep{bai2002, bai2008, Stock2006, Stock2011, LYB2011, LY2012} are among the most widely used methods for studying the temporal dynamics of high-dimensional time series. 
However, manifold-valued data are often unamenable to existing tools that are predominantly designed for data in linear spaces. 
One fundamental obstacle is that vector addition and scalar multiplication are rarely well-defined in typical nonlinear spaces. 
Moreover, because the classical factor models extract latent factors by projection onto lower-dimensional linear subspaces, they are inherently misspecified and unsuitable for analyzing intrinsically nonlinear data.

In this work, we propose a new factor model for time series lying on a Riemannian manifold, referred to as the Riemannian factor model (RFM). 
The RFM postulates that the latent factors govern the variations of the data around the Fr\'{e}chet mean. 
To estimate the RFM, we extract latent factors on the tangent space at the empirical Fr\'{e}chet mean.
Since the population and empirical Fr\'{e}chet means differ in general, the true and estimated loadings lie in different tangent spaces. Therefore, the asymptotic analysis of the RFM is more involved than classical Euclidean factor models.
%\sh{Note that since the empirical Fr\'{e}chet mean typically differs from the population Fr\'{e}chet mean, the data are nonlinearly distorted on the tangent space, so the problem does not reduce to estimating a classical factor model.} 
%{[suggest this be changed to: Since the population and empirical Fr\'{e}chet means differ in general, the true and estimated loadings lie in different tangent spaces. Therefore, the asymptotic theory for the estimation of the RFM is more involving than classical Euclidean factor models.]} 
%\sh{Our method has some advantages in prediction and interpretation because of its intrinsic formulation, while retaining the simplicity in estimation similar to the classical factor models.} 
%{[suggest this be changed to: While retaining the computational simplicity of applying classical factor models to tangent spaces, the proposed RFM provides geometry-aware prediction and enhanced interpretability by leveraging its intrinsic formulation.]}
While retaining the computational simplicity of applying classical factor models to tangent spaces, the proposed RFM provides geometry-aware prediction and enhanced interpretability by leveraging its intrinsic formulation.
For predictions, the outputs of RFM are guaranteed to lie on the manifold whereas the predictions of the linear factor models do not necessarily respect the manifold constraint. 
For interpretation, one can examine the effect of an extracted factor by studying the geodesic trajectory on the manifold traveling in the direction of the corresponding loading vector, which may offer additional insight.
In addition, we also employ a ratio estimator for selecting the number of factors. Section \ref{Sec::Setup_method} formally introduces the Riemannian factor model, discusses the related assumptions, and details the estimation procedure.

As a leading example, we model the realized covariances between selected U.S. stocks—viewed as data on the Bures--Wasserstein manifold.
Understanding the dynamics of the covariances between stocks is central in a number of financial applications, including portfolio selection \citep{Markowitz1952, YEN2014} and risk management \citep{tsay2010}.
Existing techniques \citep{Kasper2023} such as the exponential weighted moving average (EWMA), while offering reasonable estimates, provide little interpretation in what drives the changes in realized volatility. 
In contrast, our method, through factor modeling on the Bures--Wasserstein manifold, yields interpretable factors as well as competitive predictions. 
By examining the geodesics associated with the corresponding loading vectors, we found that the factors estimated by RFM capture market-wide and sector-specific effects, where the market factor closely traces the closing price of the VIX volatility index.
Section~\ref{Sec::application} contains more detail on the real data analysis. 

We establish consistency and convergence rates for the proposed RFM estimators in a high-dimensional regime where the manifold dimension is allowed to diverge with the sample size.
%For the asymptotic theory, we establish consistency and a rate of convergence of the proposed RFM estimators under the high-dimensional regime where the manifold dimension can diverge as the sample size.
In this setting, the theoretical challenges are twofold.
First, the high-dimensional regime requires a refined analysis with non-asymptotic bounds on the estimation errors, including the statistical errors and the distortion caused by the mismatch between tangent spaces of the population and the empirical Fr\'{e}chet means.
Second, the analysis should account for the temporal dependence in the data, which is further complicated by the curved geometry of the underlying manifold.
This is in sharp contrast to the existing literature on dimension reduction \citep{dai2018, Lin2019} that typically assumes the manifold dimension is fixed and data consist of independent manifold-valued measurements or functions.
Specifically, the Taylor approximation approach employed in the prior works has an unfavorable dimension dependence, and, unlike the case for functional data where independent copies of continuous sample paths are available, the temporal correlation among Riemannian time series has a direct impact on the rate of convergence.
In addition, due to the tangent space mismatch, the estimated factor loading space cannot be directly compared with the true loading space without parallel transportation.
To address these issues, we employ a novel technique by studying the holonomy, which bounds the change in the tangent vectors parallel-transported along a small geodesic triangle.
%Our convergence rates for the estimated loading space reveal a different signal-to-noise trade-off compared to the Euclidean factor models \citep{LYB2011} \yc{due to the nonlinearity of Riemannian manifolds? (add the reason of this difference and do they coincide if the manifold is flat?)}. 
In addition, under a short memory condition, a dimension-free $n^{-1/2}$ rate can still be obtained, where $n$ denotes the sample size.
A detailed discussion on the theoretical results is presented in Section \ref{Sec::theory}.

The main contributions of this work are as follows. 
First, to the best of our knowledge, this work proposes not only the first factor model but also the first dimension reduction method with theory-backed estimation for Riemannian time series.
In addition, unlike the PCA-type methods that hinge on the variance-covariance operator, our approach employs the autocovariance operator to estimate the factor loading space. 
Similar to the works of \citet{Pena1987, Pena2006, LYB2011, LY2012} for the Euclidean data, one key advantage of this approach is that it accommodates arbitrary cross-correlation in the noise, and avoids the identification issues in principal component-based factor models \citep{bai2002, bai2008, Bai2012, Forni2000, Stock2006, Stock2011}.
Second, this paper is also among the first to apply factor modeling to data on the Bures--Wasserstein manifold.
The Bures--Wasserstein metric, which is strongly motivated by optimal transport, has gained traction recently in various fields, including single-cell transcriptomics \citep{Xu2025test, Xu2025, Haviv2025}, neuroscience \citep{Zirpoli2025}, computational chemistry \citep{Jiang2025}, and diffusion tensor imaging \citep{Tang2024}. However, most prior works on dimension reduction for SPD-valued data have mostly focused on the affine-invariant \citep{Moakher2005, Pennec2006} and Log-Euclidean \citep{Arsigny2007} geometries, except for the recent work by \citet{Santoro2023} which studied functional PCA for infinite-dimensional covariance operators. 
Third, rigorous theoretical guarantees for the estimation method and the selection of the number of factors are derived under the high-dimensional Riemannian time series setup, where novel proof techniques are needed.

The paper is organized as follows. In Section \ref{Sec::Setup_method}, the RFM is formally introduced, along with an efficient estimation procedure. Section \ref{Sec::theory} provides several theoretical properties of the proposed estimators. In Section \ref{Sec::simulation}, the performance of RFM is studied using synthetically generated time series on nonlinear manifolds, including the Bures--Wasserstein manifold and the product manifold of spheres. In Section \ref{Sec::application}, we apply the RFM to the monthly covariance matrices of selected U.S.~stock returns, viewed as time series in the Bures--Wasserstein manifold. Section \ref{Sec::conclusion} concludes. All proofs to the theoretical results are presented in the supplementary material.

Next, we introduce some notations needed for the rest of the paper.
As discussed above, we adopt the high-dimensional asymptotic regime in which the manifold dimension $p = p(n)$ may diverge with the sample size $n$, following the triangular array framework of \citet{Buhlmann2011}. Hence, for example, a function $f(p, n) = f(p(n),n)$ can be shorthanded as $f(n)$. Let $(\Omega, \mathcal{F}, \mathbb{P})$ be the probability space on which the triangular array is defined. For two sequences of random variables $X_{n}, Y_{n}$, we write $X_{n} = O_{p}(Y_{n})$ if for any $\epsilon>0$ there exists $C>0$ and $N < \infty$ such that $\mathbb{P}(|X_{n}/Y_{n}| > C ) \leq \epsilon$ for all $n \geq N$. Similarly, $X_{n} = o_{p}(Y_{n})$ if for any $\epsilon > 0$, $\mathbb{P}(|X_{n}/Y_{n}| > \epsilon) \rightarrow 0$. If $X_{n}, Y_{n}$ are deterministic, we use $O$ and $o$ in lieu of $O_{p}$ and $o_{p}$, respectively. In addition, if $a_{n} = O(b_{n})$ and if $b_{n} = O(a_{n})$, we write $a_{n} \asymp b_{n}$.

\section{The Riemannian factor model: Setup and estimation} \label{Sec::Setup_method}

In this section, we introduce the Riemannian factor model (RFM), which aims to capture the dynamics of manifold-valued time series through latent factors. 
We begin by reviewing some elements of Riemannian geometry. 
Then the RFM is introduced, along with discussions about some assumptions maintained throughout the paper. 
Finally, we present an estimation procedure for RFM.

\subsection{Preliminaries}
We first review some elements of Riemannian geometry and introduce notations. 
The reader can consult, for example, \cite{docarmo1992}, for a more comprehensive treatment. 
See also \cite{Bhattacharya2003, Bhattacharya2005} and \cite{Lin2019} for brief introductions. 

Let $M$ be a connected $p$-dimensional Riemannian manifold with a Riemannian metric $\langle \cdot, \cdot \rangle_{\mathbf{x}}$, where $\mathbf{x} \in M$.
For $\mathbf{x}, \mathbf{y} \in M$, the geodesic distance is defined as $d_{M}(\mathbf{x}, \mathbf{y}) := \inf_{\gamma} \int \langle \gamma'(t), \gamma'(t) \rangle_{\gamma(t)}^{1/2} \D t$, where the infimum is over all piecewise smooth curves $\gamma$ on $M$ joining $\mathbf{x}$ and $\mathbf{y}$ and $\gamma'(t) = \D \gamma / \D t$ denotes the tangent vector field of $\gamma$.
Let $D$ be the (unique) Levi--Civita connection (see Theorem 3.6, Chapter 2, \citealp{docarmo1992}) associated with the Riemannian metric. 
A smooth curve $\gamma:[0,1] \rightarrow M$ is called a geodesic if $D_{\gamma'(t)}\gamma'(t) = 0$, meaning that its acceleration vanishes with respect to the connection. 
Geodesics are locally length-minimizing curves and serve as the analogs of straight lines in the Euclidean space. 
The tangent space of $M$ at a point $\mathbf{x}$ is denoted by $T_{\mathbf{x}}M$.
Given a tangent vector $\mathbf{v} \in T_{\mathbf{x}}M$ with sufficiently small norm $\Vert \mathbf{v} \Vert_{\mathbf{x}} = \langle \mathbf{v}, \mathbf{v} \rangle_{\mathbf{x}}^{1/2}$, there exists a unique geodesic $\gamma$ such that $\gamma(0) = \mathbf{x}$ and $\gamma'(0) = \mathbf{v}$.
For such $\mathbf{v}$, the (Riemannian) exponential map at $\mathbf{x}$ is defined as  $\rexp_{\mathbf{x}}(\mathbf{v}) = \gamma(1)$.
It is well known that there exists $R > 0$ such that $\rexp_{\mathbf{x}}$ is diffeomorphic on $V_{R} = \{\mathbf{v} \in T_{\mathbf{x}}M: \Vert \mathbf{v} \Vert < R\}$, in which case the geodesic ball with radius $R$, $B_{\mathbf{x}}(R) = \{\mathbf{y} \in M: d_{M}(\mathbf{x}, \mathbf{y}) < R\}$, is called a normal neighborhood of $\mathbf{x}$.
The supremum of such $R$'s is called the radius of injectivity. 
Furthermore, if for every $\mathbf{y} \in B_{\mathbf{x}}(R)$ there exists a normal neighborhood $G_{\mathbf{y}}$ of $\mathbf{y}$ such that $G_{\mathbf{y}} \supset B_{\mathbf{x}}(R)$, then $B_{\mathbf{x}}(R)$ is called a totally normal neighborhood.
It follows that every pair of points in the totally normal neighborhood can be connected by a unique length-minimizing geodesic.
In the rest of the paper, when there is no danger of confusion, we may omit the subscript indicating the base point $\mathbf{x} \in M$ in the Riemannian metric and its norm, and write $\langle \cdot, \cdot \rangle_{\mathbf{x}} = \langle \cdot, \cdot \rangle$, $\Vert \cdot \Vert_{\mathbf{x}} = \Vert \cdot \Vert$.

Along a geodesic $\gamma$, a vector field $V(t) \in T_{\gamma(t)}M$ is said to be parallel if $D_{\gamma'(t)}V(t) = 0$ for all $t$. 
In fact, given $V_{0} \in T_{\gamma(0)}M$, there exists a unique parallel vector field $V(t)$ along $\gamma$ with $V(0) = V_{0}$.
The tangent vector $V(t) \in T_{\gamma(t)}M$ is called the parallel-transported vector of $V_{0}$ from $\gamma(0)$ to $\gamma(t)$.
If the geodesic is unique, then by uniqueness of the parallel vector field this defines the parallel transport operator $\mathcal{P}_{\mathbf{x}\rightarrow\mathbf{y}}:T_{\mathbf{x}}M \rightarrow T_{\mathbf{y}}M$. 

The Riemannian curvature tensor $\mathcal{R}$ describes how the manifold bends. 
Formally, for vector fields $X, Y, Z$ on $M$, $\mathcal{R}(X, Y)Z = D_{X}D_{Y}Z - D_{Y}D_{X}Z - D_{[X,Y]}Z$, where $[X,Y]$ is the Lie bracket of vector fields.
Given linearly independent $\mathbf{u}, \mathbf{v} \in T_{\mathbf{x}}M$, the sectional curvature of the two-dimensional space $\sigma = \mathrm{span}\{\mathbf{u}, \mathbf{v}\}$ is defined as
\begin{align*}
    K(\mathbf{x}, \sigma) = \frac{\langle \mathcal{R}(\mathbf{u}, \mathbf{v})\mathbf{u}, \mathbf{v} \rangle}{\Vert \mathbf{u} \Vert^{2}\Vert \mathbf{v} \Vert^{2} - \langle \mathbf{u}, \mathbf{v} \rangle^2}.
\end{align*}
The definition is independent of the choice of the basis $\{\mathbf{u}, \mathbf{v}\}$ for $\sigma$.
Nonzero curvature fundamentally distinguishes a curved space from a flat one, such as the Euclidean space.
Hence, it is crucial in both the study of Riemannian geometry and our theoretical analysis.

As a concrete example, the set of real symmetric positive-definite (SPD) matrices of the same dimensions, equipped with the Bures--Wasserstein metric, forms a Riemannian manifold \citep{Malago2018, Chewi2025, Takatsu2011}.
%Details of the metric can be found in \textcolor{red}{Section S.2 of the supplementary material}.
The resulting geodesic distance between two SPD matrices $\mathbf{\Sigma}_{1}$, $\mathbf{\Sigma}_{2}$ is $\{\mathrm{Tr}(\mathbf{\Sigma}_{1} + \mathbf{\Sigma}_{2} - 2 (\mathbf{\Sigma}_{1}^{1/2}\mathbf{\Sigma}_{2}\mathbf{\Sigma}_{1}^{1/2})^{1/2})\}^{1/2}$.
This corresponds to the 2-Wasserstein distance between the two normal distributions $N(0, \mathbf{\Sigma}_{1})$ and $N(0, \mathbf{\Sigma}_{2})$, which is the optimal transport cost for transferring mass from the distribution $N(0, \mathbf{\Sigma}_{1})$ to $N(0, \mathbf{\Sigma}_{2})$. 
Due to this connection to optimal transport, the Bures--Wasserstein metric appears to be more natural to econometric and financial applications compared to other Riemannian metrics, such as the affine-invariant \citep{Moakher2005, Pennec2006} or the log-Euclidean \citep{Arsigny2007} metrics---in fact, optimal transport models are widely seen in economics and finance.
For example, in portfolio selection, \citet{Blanchet2022} employ a distributionally robust approach where the ambiguity set is a Wasserstein ball (see also \citealp{Pesenti2023}).
As another example, model-independent hedging is related to the martingale optimal transport problem \citep{DeMarco2015, Henry-Labordere2017}.
\cite{galichon2016optimal} provides an excellent overview of the applications of optimal transport in econometrics. 
In Section S.2 of the supplementary material, we briefly review the properties of the Bures--Wasserstein manifold.
However, contrary to the affine-invariant and the log-Euclidean metrics, the Bures--Wasserstein manifold has nonnegative sectional curvature and is incomplete. 
Although these properties introduce some complexity to the theoretical analysis, our theoretical results are applicable in this setting; see Section \ref{Sec::theory} for further discussions.

\subsection{The Riemannian factor model}

We now introduce the Riemannian factor model (RFM) for time series taking values in a Riemannian manifold $M$. 
Let $\mathbf{x}_{t} \in M$ be the observed manifold-valued time series. 
The RFM assumes 
\begin{align}
    \mathbf{x}_{t} &= \rexp_{\bm{\mu}}(\mathbf{z}_{t}), \label{new1} 
\end{align}
where $\bm{\mu} \in M$ is a fixed reference point and the latent time series $\mathbf{z}_{t}$ is mean-zero and takes values in the $p$-dimensional tangent space $T_{\bm{\mu}}M$. 
Under some assumptions, $\bm{\mu}$ can be viewed as the Fr\'{e}chet mean of $\mathbf{x}_t$ (See Remark \ref{rmk:mu} below).
Intuitively, the latent time series $\mathbf{z}_{t}$ represents the deviation from $\bm{\mu}$, which is mapped to the observed $\mathbf{x}_{t}$ via the Riemannian exponential.
Then RFM specifies a factor model for $\mathbf{z}_t$ in the tangent space,
\begin{align}
    \mathbf{z}_{t} &= \mathcal{A}(\mathbf{f}_{t}) + \bm{\delta}_{t}, \label{new2}
\end{align}
where $\mathbf{f}_{t} \in \mathbb{R}^{r}$ with $r \ll p$, is the factor process, $\bm{\delta}_{t}$ is white noise, and $\mathcal{A}:\mathbb{R}^{r} \rightarrow T_{\bm{\mu}}M$ is an isometric linear operator. 
Hence $\mathbf{z}_{t}$ is governed by only a few latent factors.
Here we adopt the framework of \cite{LYB2011} and \cite{LY2012} where the factors account for the dynamic correlations, and it is assumed that no linear combination of $\mathbf{f}_{t}$ would result in a white noise. 
This allows the covariance of $\bm{\delta}_{t}$ to be arbitrary.
Different from the PCA or PGA models adopted in the existing non-Euclidean statistics literature \citep{Fletcher2004, Fletcher2007, Huckemann2010, dai2018, Lin2019}, our focus is on the dynamic correlations.

To illustrate the factor model formulation \eqref{new2}, fix an orthonormal basis $\mathbf{E} = (\mathbf{e}_{1}, \ldots, \mathbf{e}_{p})$ of $T_{\bm{\mu}}M$. 
We can express $\mathbf{z}_{t}$ as a vector $\mathbf{z}_{t, \mathbf{E}} \in \mathbb{R}^{p}$ with respect to $\mathbf{E}$.
The linear factor model of \citet{LYB2011} assumes
\begin{align} \label{Sec2-RFM1}
    \mathbf{z}_{t, \mathbf{E}} = \mathbf{A}_{\mathbf{E}} \mathbf{f}_{t} + \bm{\delta}_{t, \mathbf{E}},
\end{align}
where $\mathbf{A}_{\mathbf{E}}$ is assumed to have orthonormal columns without loss of generality. 
The following proposition shows the two models \eqref{new2} and \eqref{Sec2-RFM1} are equivalent.
\begin{prop} \label{prop:model}
    If \eqref{new2} holds with
    \begin{enumerate}[label=(F)] 
        \item\label{F_intrinsic} $\{\bm{\delta}_{t}\}$ is a mean-zero white noise, and $\mathcal{A}$ is isometric,
    \end{enumerate}
    then for any orthonormal basis $\mathbf{E}$ of $T_{\bm{\mu}}M$, $\mathbf{z}_{t, \mathbf{E}}$ satisfies \eqref{Sec2-RFM1} with
    \begin{enumerate}[label=(F')]
        \item\label{F_coordinate} $\bm{\delta}_{t,\mathbf{E}} \in \mathbb{R}^{p}$ is a mean-zero white noise, and $\mathbf{A}_{\mathbf{E}}$ has orthonormal columns.
    \end{enumerate}
    Conversely, if \eqref{Sec2-RFM1} holds with \ref{F_coordinate} for some basis $\mathbf{E}$, then \eqref{new2} holds with \ref{F_intrinsic}. 
\end{prop}

The factor representation \eqref{Sec2-RFM1} is basis-dependent.
If a different set of orthonormal basis $\mathbf{D}$ is used, then there exists an unitary $\mathbf{Q} \in \mathbb{R}^{p \times p}$ such that $\mathbf{z}_{t, \mathbf{D}} = \mathbf{Q} \mathbf{z}_{t, \mathbf{E}} = \mathbf{QA}_{\mathbf{E}} \mathbf{f}_{t} + \mathbf{Q} \bm{\delta}_{t,\mathbf{E}}$, and $\mathbf{z}_{t,\mathbf{D}}$ still admits a factor model structure by Proposition \ref{prop:model}.
In practice, we often need to use a specific basis for computation (see the next subsection), but it is easier to work with the intrinsic formulation \eqref{new2}, which avoids the ambiguity of the choice of the basis, in the theoretical expositions.

We make some assumptions about the RFM and discuss here their implications. 
First, as noted above, we assume $\bm{\delta}_{t}$ is a white noise while $\mathbf{f}_{t}$ drives all dynamics. Specifically, we assume
$\{\mathbf{f}_{t}\}$ and $\{\bm{\delta}_{t}\}$ are weakly stationary processes such that $\mathbb{E}(\bm{\delta}_{t} \otimes \bm{\delta}_{t+h}) = 0$ and $\mathbb{E}(\bm{\delta}_{t+h} \otimes \mathbf{f}_{t}) = 0$ for $h \neq 0$, where for two vectors $\mathbf{u}, \mathbf{v}$ from the vector spaces $U, V$, respectively, the operator $\mathbf{u} \otimes \mathbf{v}: U \rightarrow V$ is defined by $(\mathbf{u} \otimes \mathbf{v})\mathbf{w} = \langle \mathbf{u}, \mathbf{w} \rangle \mathbf{v}$ for all $\mathbf{w} \in U$.
Second, let $Q_{t}$ be the probability measure of $\mathbf{x}_{t}$.
The Fr\'{e}chet mean of $Q_{t}$ is defined as the minimizer of
\begin{align*}
    F_{t}(\mathbf{y}) = \int d_{M}(\mathbf{y}, \mathbf{x})^{2} Q_{t}(\D\mathbf{x}), \quad \mathbf{y} \in M,
\end{align*}
provided the minimizer exists and is unique. 
We assume the following about $\{\mathbf{x}_{t}\}$.
\begin{enumerate}[label=(P\arabic*),series=P]
    \item\label{ass:support} There exists $R > 0$ such that the geodesic ball around $\bm{\mu}$ with radius $R$, $B_{\bm{\mu}}(R) = \{\mathbf{x} \in M: d_{M}(\mathbf{x}, \bm{\mu}) < R \}$, is a normal neighborhood of $\bm{\mu}$, and there exists some small $\eta > 0$ such that $\mathbb{P}(\mathbf{x}_{t} \in B_{\bm{\mu}}(R/(2 + \eta))) = 1$ for all $t$. 
    \item\label{ass:constant_mean} The Fr\'{e}chet mean of $Q_{t}$ exists and is equal to $\bm{\mu}$ for all $t$.
\end{enumerate}

\ref{ass:support} assumes the data cluster around $\bm{\mu}$, which implies the geodesic joining $\mathbf{x}_{t}$ and $\bm{\mu}$ varies smoothly and that closed subsets in this neighborhood are compact.
For spheres, $B_{\bm{\mu}}(R)$ is a normal neighborhood if $R \leq \pi$.
For negatively curved manifolds, such as the SPD matrices equipped the affine-invariant metric, exponential maps are global diffeomorphisms on these manifolds. 
It follows that geodesic balls around $\bm{\mu}$ are totally normal neighborhoods, and \ref{ass:support} holds as long as the data are within bounded distance from $\bm{\mu}$.
%The first part of \ref{ass:constant_mean} assumes the Fr\'{e}chet mean of $Q_{t}$ exists. 
By \eqref{new1}, \ref{ass:support}, and an argument similar to Theorem 2.1 of \cite{Bhattacharya2003}, the set of minimizers of $F_{t}$ over $M$ is nonempty. 
Thus, for the existence of the Fr\'{e}chet mean \ref{ass:constant_mean} essentially assumes uniqueness of the minimizer.
In general, uniqueness is guaranteed if $Q_{t}$ is sufficiently concentrated, depending on the curvature of the underlying manifold (see Remark 2.1 of \citealp{Bhattacharya2003}).

\begin{remark} \label{rmk:mu}
    %Here we provide an interpretation for $\bm{\mu}$ in \eqref{new1}. 
    The second part of \ref{ass:constant_mean} requires the Fr\'{e}chet means of $\mathbf{x}_{t}$, $t=1,2,\ldots$, are equal to $\bm{\mu}$.
    From a modeling perspective, this assumption is quite natural.
    Indeed, if the Fr\'{e}chet mean of $Q_{t}$ is $\bm{\nu}$ and if $\mathbb{P}(\mathbf{x}_{t} \in B_{\bm{\nu}}(R_{\bm{\nu}})) = 1$, where $R_{\bm{\nu}}$ is the radius of injectivity at $\bm{\nu}$, then it follows from Theorem 2.1 of \cite{Bhattacharya2003} that
    \begin{align*}
        \int \rlog_{\bm{\nu}}(\mathbf{x}) Q_{t}(\D\mathbf{x}) = 0,
    \end{align*}
    where $\rlog_{\bm{\nu}}: B_{\bm{\nu}}(R_{\bm{\nu}}) \rightarrow T_{\bm{\nu}}M$ is the inverse of the exponential map at $\bm{\nu}$, restricted to the ball centered at zero with the radius $R_{\bm{\nu}}$.
    Then Eq.~\eqref{new2} simply postulates a factor model for the mean-zero process $\mathbf{z}_{t} = \rlog_{\bm{\nu}}(\mathbf{x}_{t})$ in the tangent space.
\end{remark}

In high-dimensional statistics, concentration inequalities are indispensable tools \citep{Wainwright_2019, Buhlmann2011, Boucheron}. For RFM, we assume the following. 

\begin{enumerate}[label=(P\arabic*),resume=P]
    \item\label{ass:concentration} $\{\mathbf{z}_{t}\}$ satisfies the concentration inequality
    \begin{align*}
        \sup_{\mathbf{w} \in B_{\bm{\mu}}(R)} \left\vert \frac{1}{n} \sum_{t=1}^{n} [d_{M}(\mathbf{w}, \rexp_{\bm{\mu}}(\mathbf{z}_{t}))^{2} - \mathbb{E}(d_{M}(\mathbf{w}, \rexp_{\bm{\mu}}(\mathbf{z}_{t}))^{2}) ]\right\vert = O_{p}(c_{n}),
    \end{align*}
    where $c_{n} \rightarrow 0$.
\end{enumerate}

In general, \ref{ass:concentration} is a special case of the concentration inequality 
\begin{align*}
        \sup_{f \in \mathcal{F}_{n}} \left\vert \frac{1}{n}\sum_{t=1}^{n} [f(\mathbf{z}_{t}) - \mathbb{E}f(\mathbf{z}_{t})] \right\vert = O_{p}(c_{n}),
    \end{align*}
where $f: T_{\bm{\mu}}M \rightarrow \mathbb{R}$ and $\mathcal{F}_{n}$ is some suitable class of functions.
Such concentration inequalities hold for a wide variety of time series processes. The concentration rate $c_{n}$ typically hinges on the temporal dependence structure and the complexity of the function class $\mathcal{F}_{n}$.
For instance, \cite{Chen2018} and \cite{deb2024} established concentration inequalities under various dependence assumptions characterized by the rate of decay in moving average coefficients or mixing conditions, both of which allow long-range dependent and heavy-tailed processes.
See also \cite{Yu1994} for a uniform bound characterized by the metric entropy of $\mathcal{F}_{n}$ and the decay rate of mixing coefficients.
Note that the dependence on the manifold dimension $p$ is absorbed into the rate $c_{n}$. 
Below we give an example where $c_{n}$ is explicitly calculated.

\begin{example} \label{eg:c_n}
    Suppose \ref{ass:support} holds and $M$ is complete with sectional curvature bounded below by $K_{1} \in \mathbb{R}$. Assume $\{\mathbf{z}_{t}\}$ is a strictly stationary $\alpha$-mixing process with $\alpha$-mixing coefficients $\{\alpha_{j}\}$. Then
    \begin{itemize}
        \item (Geometric decay) If $\{\alpha_{k}\}$ satisfies $\alpha_{k} \leq H \rho^{k}$ for some $H > 0$, $\rho \in (0, 1)$ and if $p = o(n^{\gamma}/\log n)$ for some $\gamma \in (0, 0.5)$, then \ref{ass:concentration} holds with $c_{n} = n^{-(1 - 2\gamma)/2}$.
        \item (Algebraic decay) If $\alpha_{k} \leq H k^{-\zeta}$ for some $H>0$ and $\zeta > 0$, and if $p$ is fixed, then \ref{ass:concentration} holds with $c_{n} = n^{-\frac{\zeta}{3/2 + p + 2\zeta}} \log n$.
    \end{itemize}
\end{example}

%The proof of Example \ref{eg:c_n} is presented in Appendix \ref{App::theory}. 
This example indicates a trade-off between the serial dependence in the data and the manifold dimension $p$.
If $\mathbf{z}_{t}$ has geometrically decaying $\alpha$-mixing coefficients, \ref{ass:concentration} holds with $p$ diverging at a rate slower than $\sqrt{n}$. 
If the $\alpha$-mixing coefficients decay algebraically, a sufficient condition for \ref{ass:concentration} is that $p$ is fixed, with the convergence rate $c_{n}$ adversely affected by a large $p$.
The faster the $\alpha$-mixing coefficients tend to zero (a large $\zeta$), the faster the convergence rate $c_{n}$ is, which is close to the ideal $n^{-1/2}$ rate in the limiting case where $\zeta \rightarrow \infty$.
In Section S.1 of the supplementary material, we also discuss the $\alpha$-mixing coefficients of high-dimensional linear processes, which may be of independent interest. 
In particular, we show that for a high-dimensional VAR(1) process, the $\alpha$-mixing coefficients decay geometrically but the constant $H$ in Example \ref{eg:c_n} can scale polynomially with $p$.
Nevertheless, the result of Example \ref{eg:c_n} still holds for such process because $H$ only has a logarithmic effect (see Example S.1 and Remark S.3 in the supplementary material).

For the manifold $M$, we assume 
\begin{enumerate}[label=(M\arabic*),series=M]
    \item\label{ass:curvature} The sectional curvature satisfies $K_{1} \leq K(\mathbf{x}, \sigma) \leq K_{0}$ for  all $\mathbf{x} \in B_{\bm{\mu}}(R)$ and all two-dimensional subspace $\sigma \subset T_{\mathbf{x}}M$, for some $K_{0} < \infty$ and $K_{1} > -\infty$.
\end{enumerate}
For many standard manifolds, either $K_{0}$ or $K_{1}$ is trivial.
For example, the sphere or the Bures–Wasserstein manifold are nonnegatively curved, i.e., $K(\mathbf{x}, \sigma) \geq 0$, while the space of SPD matrices equipped with the affine-invariant metric satisfies $K(\mathbf{x}, \sigma) \leq 0$.

In closing the subsection, we remark that the Riemannian factor model can be employed for {\it multivariate} manifold-valued data.
For instance, if $\mathbf{x}_{t}$ and $\mathbf{y}_{t}$ are time series taking values in the Riemannian manifolds $M_{1}$ and $M_{2}$ respectively, then one can view $(\mathbf{x}_{t}, \mathbf{y}_{t})$ as a time series in the product manifold $M_{1} \times M_{2}$, which inherits a Riemannian structure.
In this setting, the RFM captures the lower-dimensional factors driving the joint dynamic dependence of $\mathbf{x}_{t}$ and $\mathbf{y}_{t}$.
In Section \ref{Sec::simulation}, we investigate the application of RFM to the product manifold of spheres in the simulation study.

\subsection{Estimation}

Let $\mathbf{x}_{1}, \mathbf{x}_{2}, \ldots, \mathbf{x}_{n} \in M$ be the observed time series. We estimate the RFM as follows.
First, the Fr\'{e}chet mean is estimated by optimizing
\begin{align*}
    \hat{\bm{\mu}} \in \arg\min_{\mathbf{y} \in M} \sum_{t=1}^{n} d_{M}(\mathbf{y}, \mathbf{x}_{t})^{2}.
\end{align*}
Numerically, Riemannian gradient methods are often employed to solve this minimization problem (see, for example, \citealp{Fletcher2004}).
Let $\hat{\mathbf{E}} = (\hat{\mathbf{e}}_{1}, \ldots, \hat{\mathbf{e}}_{p})$ be an orthonormal basis of $T_{\hat{\bm{\mu}}}M$.
Then we can express the log-mapped data $\hat{\mathbf{z}}_{t} = \rlog_{\hat{\bm{\mu}}} (\mathbf{x}_{t})$ with respect to $\hat{\mathbf{E}}$, denoted as $\hat{\mathbf{z}}_{t, \hat{\mathbf{E}}}$. 
Let $\hat{\mathbf{S}}_{\hat{\mathbf{E}}}(h) = n^{-1} \sum_{t = h_{0}+1}^{n} \hat{\mathbf{z}}_{t, \hat{\mathbf{E}}} \hat{\mathbf{z}}_{t-h, \hat{\mathbf{E}}}^{\top}$, $h = 1, 2, \ldots, h_{0}$, be the sample autocovariance matrices expressed in $\hat{\mathbf{E}}$, where $h_{0} \geq 1$ is a prescribed positive integer.   
Following \cite{LYB2011}, put $\hat{\mathbf{L}}_{\hat{\mathbf{E}}} = \sum_{h=1}^{h_{0}} \hat{\mathbf{S}}_{\hat{\mathbf{E}}}(h) \hat{\mathbf{S}}_{\hat{\mathbf{E}}}(h)^{\top}$ and compute the first $r$ eigenvectors $\hat{\mathbf{A}}_{\hat{\mathbf{E}}} = (\hat{\mathbf{a}}_{1,\hat{\mathbf{E}}}, \ldots, \hat{\mathbf{a}}_{r,\hat{\mathbf{E}}})$ corresponding the largest $r$ eigenvalues of $\hat{\mathbf{L}}_{\hat{\mathbf{E}}}$. 
Subsequently, the factor process can be estimated by $\hat{\mathbf{f}}_{t} = \hat{\mathbf{A}}_{\hat{\mathbf{E}}}^{\top} \hat{\mathbf{z}}_{t, \hat{\mathbf{E}}}$.
Let $J_{\hat{\mathbf{E}}}: \mathbb{R}^{p} \rightarrow T_{\hat{\bm{\mu}}}M$ be defined as
\begin{align} \label{J_def}
    J_{\hat{\mathbf{E}}}(a_{1}, \ldots, a_{p}) = \sum_{j=1}^{p}a_{j}\hat{\mathbf{e}}_{j}.
\end{align} 
That is, $J_{\hat{\mathbf{E}}}$ identifies the $\hat{\mathbf{E}}$-coordinate vector with the corresponding element in the tangent space. 
Then the (in-sample) model predictions are $\hat{\mathbf{x}}_{t} = \rexp_{\hat{\bm{\mu}}}(J_{\hat{\mathbf{E}}}(\hat{\mathbf{A}}_{\hat{\mathbf{E}}}\hat{\mathbf{f}}_{t}))$.
In practice, the number of factors $r$ is unknown. 
We may estimate the number of factors by
\begin{align} \label{Sec2-ratio}
    \hat{r} = \arg\min_{1 \leq i \leq R} \hat{\lambda}_{i+1} / \hat{\lambda}_{i},
\end{align}
where $\hat{\lambda}_{1} \geq \hat{\lambda}_{2} \geq \ldots$ are the eigenvalues of $\hat{\mathbf{L}}_{\hat{\mathbf{E}}}$.
For Euclidean data, this ratio estimator is employed by \citet{LY2012}.

Due to the intrinsic formulation of the Riemannian factor model, the evaluation of the accuracy and performance requires special consideration.
Our focus is on the uniquely defined loading space in the factor model \eqref{new2}, which is the image space of $\mathcal{A}$, $\mathrm{Im}(\mathcal{A})$. 
To avoid the dependence on the choice of the basis, we shall refer some key quantities above to their intrinsic counterparts. Specifically, for $\hat{\mathbf{S}}_{\hat{\mathbf{E}}}$ and $\hat{\mathbf{L}}_{\hat{\mathbf{E}}}$, their intrinsic, basis-independent versions are, respectively,
\begin{align*}
    \hat{\mathcal{S}}(h) = \frac{1}{n} \sum_{t=h_{0}+1}^{n} \hat{\mathbf{z}}_{t-h} \otimes \hat{\mathbf{z}}_{t}, \mbox{ and}\quad \hat{\mathcal{L}} = \sum_{h=1}^{h_{0}} \hat{\mathcal{S}}(h) \hat{\mathcal{S}}(h)^{*},
\end{align*}
where $\mathcal{A}^{*}$ denotes the adjoint of the operator $\mathcal{A}$.
Let $\hat{\mathbf{a}}_{1}, \ldots, \hat{\mathbf{a}}_{r} \in T_{\hat{\bm{\mu}}}M$ be the leading $r$ eigenvectors of $\hat{\mathcal{L}}$, which spans the estimated loading space $\mathrm{span}\{\hat{\mathbf{a}}_{1}, \ldots, \hat{\mathbf{a}}_{r}\}$.
Note that $\{\hat{\lambda}_{i} : i=1,2,\ldots\}$ in Eq.~\eqref{Sec2-ratio} are also eigenvalues of $\hat{\mathcal{L}}$.

Note that $\hat{\mathcal{S}}(h)$, $\hat{\mathcal{L}}$, and $\{\hat{\mathbf{a}}_{j}\}$ are associated with $T_{\hat{\bm{\mu}}}M$, which are not directly comparable to the counterparts acting on $T_{\bm{\mu}}M$.
To tackle this issue in theoretical investigation and in simulation studies, we parallel transport the estimated loading space to a subspace in $T_{\bm{\mu}}M$ along the unique geodesic between $\hat{\bm{\mu}}$ and $\bm{\mu}$, and compare the two subspaces using the sine-theta distance. In particular, we measure 
\begin{align} \label{sinetheta_def}
    \Vert \sin \Theta(\mathrm{span}\{\hat{\mathcal{P}}_{\hat{\bm{\mu}} \rightarrow \bm{\mu}} \hat{\mathbf{a}}_{1}, \ldots, \hat{\mathcal{P}}_{\hat{\bm{\mu}} \rightarrow \bm{\mu}}\hat{\mathbf{a}}_{r}\}, \mathrm{Im}(\mathcal{A})) \Vert_{op},
\end{align}
where $\sin \Theta(\mathcal{W}, \mathcal{V})$ is the diagonal matrix of the sine-theta angles between the two linear spaces $\mathcal{W}, \mathcal{V} \subset T_{\bm{\mu}}M$ \citep{davis1970, chen2021}, and $\Vert \cdot \Vert_{op}$ denotes the operator norm.
This strategy, also employed by \cite{Lin2019}, involves only intrinsic quantities and therefore can be evaluated without recourse to an ambient space in which the manifold $M$ is embedded. 
For manifolds like the Bures--Wasserstein manifold which lacks a canonical ambient space, this is particularly useful.

\begin{remark}
    Following \cite{Lin2019}, a linear operator $\hat{\mathcal{B}}:T_{\hat{\bm{\mu}}}M \rightarrow T_{\hat{\bm{\mu}}}M$ can be compared with the linear operator $\mathcal{B}:T_{\bm{\mu}}M \rightarrow T_{\bm{\mu}}M$ through the parallel transporter $\Phi_{\hat{\bm{\mu}}, \bm{\mu}}$, defined as
    \begin{align} \label{Phi_def}
        (\Phi_{\hat{\bm{\mu}}, \bm{\mu}} \hat{\mathcal{B}})\mathbf{v} = \mathcal{P}_{\hat{\bm{\mu}} \rightarrow \bm{\mu}} \hat{\mathcal{B}} (\mathcal{P}_{\bm{\mu} \rightarrow \hat{\bm{\mu}}} \mathbf{v}) \quad \mbox{for all } \mathbf{v} \in T_{\bm{\mu}}M.
    \end{align}
    Let $\mathcal{B} = \mathcal{A}\mathcal{A}^{*}$ and $\hat{\mathcal{B}} = \sum_{j=1}^{r} \hat{\mathbf{a}}_{j} \otimes \hat{\mathbf{a}}_{j}$.
    Since $\mathrm{Im}(\mathcal{B}) = \mathrm{Im}(\mathcal{A})$ and $\mathrm{Im}(\Phi_{\hat{\bm{\mu}}, \bm{\mu}}\hat{\mathcal{B}}) = \newline \mathrm{span}\{\mathcal{P}_{\hat{\bm{\mu}} \rightarrow \bm{\mu}}\hat{\mathbf{a}}_{1}, \ldots, \mathcal{P}_{\hat{\bm{\mu}} \rightarrow \bm{\mu}}\hat{\mathbf{a}}_{r}\}$, the sine-theta distance 
    \begin{align*}
        \Vert \sin \Theta(\mathrm{span}\{\hat{\mathcal{P}}_{\hat{\bm{\mu}} \rightarrow \bm{\mu}} \mathbf{a}_{1}, \ldots, \hat{\mathcal{P}}_{\hat{\bm{\mu}} \rightarrow \bm{\mu}}\hat{\mathbf{a}}_{r}\}, \mathrm{Im}(\mathcal{A})) \Vert_{op}
    \end{align*}
    %$\Vert \sin \Theta(\mathrm{span}\{\hat{\mathcal{P}}_{\hat{\bm{\mu}} \rightarrow \bm{\mu}} \mathbf{a}_{1}, \ldots, \hat{\mathcal{P}}_{\hat{\bm{\mu}} \rightarrow \bm{\mu}}\hat{\mathbf{a}}_{r}\}, \mathrm{Im}(\mathcal{A})) \Vert_{op}$ 
    can be interpreted as the discrepancy between the operators $\mathcal{B}$ and $\hat{\mathcal{B}}$ under parallel transporter. 
\end{remark}

\section{Theoretical results} \label{Sec::theory}
In this section, we present the theoretical properties of the proposed estimators.
We first establish a convergence rate of the estimators under fairly mild assumptions.
Then, we show that with a short-memory and a strong factor assumption, a dimension-free convergence rate can be attained.
All proofs are relegated to the supplementary material.

To begin, we show that the estimator $\hat{\bm{\mu}}$ for the Fr\'{e}chet mean $\bm{\mu}$ is consistent.
For this, we assume the following identification conditions.

\begin{enumerate}[label=(P\arabic*),resume=P]
    \item\label{ass:loss_bound_around_mean} 
    For any $\epsilon > 0$,
    \begin{align*}
        \inf_{n \in \mathbb{N}} \inf_{t \in [n]} \inf_{\mathbf{w}:d_{M}(\mathbf{w}, \bm{\mu})>\epsilon} \{\mathbb{E}(d_{M}(\mathbf{w}, \mathbf{x}_{t})^2) - \mathbb{E}(d_{M}(\bm{\mu}, \mathbf{x}_{t})^2)\} > 0.
    \end{align*}
    %For any $\epsilon > 0$,
    %\begin{align*}
    %    \inf_{n \in \mathbb{N}} \inf_{\mathbf{w}:d_{M}(\mathbf{w}, \bm{\mu})>\epsilon} {\{}\mathbb{E}(d_{M}(\mathbf{w}, \mathbf{x}_{t})^2) - \mathbb{E}(d_{M}(\bm{\mu}, \mathbf{x}_{t})^2) {\}}> 0.
    %\end{align*}
    \item\label{ass:hessian} 
    Let $g_{n}(\mathbf{v}) = n^{-1} \sum_{t=1}^{n} \int d_{M}(\rexp_{\bm{\mu}}(\mathbf{v}), \mathbf{x})^{2} Q_{t}(\D\mathbf{x})$. There exists some $B>0$ such that
    \begin{align*}
        \sup_{\Vert \mathbf{v} \Vert < B} \lambda_{\min}^{-1} \left( \frac{\partial^{2}}{\partial \mathbf{v}^{2}} g_{n}(\mathbf{v}) \right) = O(1).
    \end{align*}
    %{Let $Q\equiv Q_t$ due to stationarity condition} {where?} {see my suggestion in Remark~3; refer to the condition after listing it as an assumption (Px)} {strict stationarity is removed everywhere} %, and $g_{n}(\mathbf{v}) = \int d_{M}(\rexp_{\bm{\mu}}(\mathbf{v}), \mathbf{x})^{2} Q(\D\mathbf{x})$. There exists some $B>0$ such that
    %\begin{align*}
    %    \sup_{\Vert \mathbf{v} \Vert < B} \lambda_{\min}^{-1} \left( \frac{\partial^{2}}{\partial \mathbf{v}^{2}} g_{n}(\mathbf{v}) \right) = O(1),
    %\end{align*}
    %where $\lambda_{\min}(\mathbf{A})$ {$(\mathbf{A})$ is missing in the displayed equation} denotes the minimum eigenvalue of a matrix or operator $\mathbf{A}$. 
\end{enumerate}

\ref{ass:loss_bound_around_mean} and \ref{ass:hessian} assume the loss function $n^{-1} \sum_{t=1}^{n} \int d_{M}(\cdot, \mathbf{x})^{2} Q_{t}(\D \mathbf{x})$ %where $Q_{t} \equiv Q$ because of stationarity, 
is well-behaved around the minimizer $\bm{\mu}$.
Similar assumptions were also adopted in \cite{dai2018} and \cite{Lin2019}.

\begin{prop} \label{prop1}
    Assume \ref{ass:support}--\ref{ass:loss_bound_around_mean}. Then
    \begin{align} \label{prop1-statement1}
        d_{M}(\hat{\bm{\mu}}, \bm{\mu}) = o_{p}(1).
    \end{align}
    Moreover, if \ref{ass:support}--\ref{ass:hessian} holds, then
    \begin{align} \label{prop1-statement2}
        d_{M}(\hat{\bm{\mu}}, \bm{\mu}) = O_{p}(c_{n}^{1/2}),
    \end{align}
    where $c_n$ is given in assumption \ref{ass:concentration}.
\end{prop}

%\begin{prop} \label{prop2}
%    Assume \ref{ass:support}--\ref{ass:hessian}. Then 
%    \begin{align*}
%        d_{M}(\hat{\bm{\mu}}, \bm{\mu}) = O_{p}(c_{n}^{1/2}). 
%    \end{align*}
%\end{prop}

%In the following, let $\mathcal{W}, \mathcal{V}$ be two $r$-dimensional linear subspaces of $T_{\bm{\mu}}M$ and let $\{\mathbf{w}_{1}, \ldots, \mathbf{w}_{r}\}$, $\{\mathbf{v}_{1}, \ldots, \mathbf{v}_{r}\}$ be the orthonormal basis, respectively, for $\mathcal{W}$ and $\mathcal{V}$, expressed in a common orthonormal basis of $T_{\bm{\mu}}M$. 

With the convergence rate of the sample Fr\'{e}chet mean $\hat{\bm\mu}$ to the Fr\'{e}chet mean $\bm\mu$ in Proposition~\ref{prop1}, next we establish a convergence rate for the estimated loading space, which is given in Theorem \ref{prop4} below.

\begin{theorem} \label{prop4}
Assume \ref{ass:curvature}, \ref{ass:support}--\ref{ass:hessian}, and that there exists $\delta\in(0,\eta)$ with $\eta$ defined in Assumption~\ref{ass:support} such that the radius of injectivity of every point of $B_{\bm{\mu}}(R/(2+\eta))$ is at least $(1+\delta)R/(2+\eta)$. If $R < (\frac{2+\eta}{1+\delta}) \frac{\pi}{\sqrt{K_{0}}}$ with $K_0$ given in Assumption \ref{ass:curvature}, then
%Assume \ref{ass:curvature}, \ref{ass:support}--\ref{ass:hessian} with $B_{\bm{\mu}}(R/(2+\eta))$ being a totally normal neighborhood and $R < \pi / ((2+ \epsilon)\sqrt{K_{0}})$ for some $\epsilon>0$. Then,
\begin{align} \label{prop4-statement-1}
    \Vert \Phi_{\hat{\bm{\mu}}, \bm{\mu}} \hat{\mathcal{L}} - \mathcal{L} \Vert_{op} = O_{p}(c_{n}^{1/2}),
\end{align}
where $\Phi_{\hat{\bm{\mu}}, \bm{\mu}}$ is defined in Eq.~\eqref{Phi_def}, $\mathcal{L} = \sum_{h=1}^{h_{0}} \mathcal{S}(h) \mathcal{S}(h)^{*}$ and $\mathcal{S}(h) = n^{-1} \sum_{t = h_{0}+1}^{n} \mathbf{z}_{t-h} \otimes \mathbf{z}_{t}$, and $c_n$ is as in Assumption \ref{ass:concentration}.
In addition, if for any fixed positive integer $h$,
\begin{align} \label{P3-2}
    \left\Vert \frac{1}{n}\sum_{t=h+1}^{n} (\mathbf{z}_{t} \otimes \mathbf{z}_{t-h} - \mathbb{E}(\mathbf{z}_{t} \otimes \mathbf{z}_{t-h})) \right\Vert_{\mathrm{HS}} = O_{p}(\epsilon_{n}),
\end{align}
where $\epsilon_{n} = o(1)$ and $\Vert \cdot \Vert_{\mathrm{HS}}$ denotes the Hilbert--Schmidt norm, then the space spanned by the first $r$ eigenvectors, $\hat{\mathbf{a}}_{1}, \ldots, \hat{\mathbf{a}}_{r}$, of $\hat{\mathcal{L}}$ satisfies
\begin{align} \label{prop4-statement-2}
    \left\Vert \sin \Theta( \mathrm{span}\{\mathcal{P}_{\hat{\bm{\mu}} \rightarrow \bm{\mu}} \hat{\mathbf{a}}_{1} \ldots \mathcal{P}_{\hat{\bm{\mu}} \rightarrow \bm{\mu}} \hat{\mathbf{a}}_{r}\}, \mathrm{Im}(\mathcal{A}) ) \right\Vert_{op} = O_{p}\left(\frac{\sqrt{c_{n}}+\epsilon_{n}}{\kappa^{2}}\right),
\end{align}
where $r$ is the dimension of the factor process $\mathbf{f}_t$ as defined after Eq.~\eqref{new2}, $\sin\Theta(\cdot,\cdot)$ is defined in Eq.~\eqref{sinetheta_def}, and $\kappa = \max_{1 \leq h \leq h_{0}} \sigma_{r}(\mathbb{E}(\mathbf{f}_{t-h} \otimes \mathbf{f}_{t}))$ with $\sigma_{k}(\cdot)$ denoting the $k$-th singular value (in descending order).
%, and $\sin \Theta(\mathcal{W}, \mathcal{V})$ is the diagonal matrix of sine-theta angles between the two linear spaces $\mathcal{W}, \mathcal{V} \subset T_{\bm{\mu}}M$ (\citealp{davis1970, chen2021}).
\end{theorem}

Theorem \ref{prop4} is novel in two aspects.
First, unlike the convergence rates obtained in \cite{dai2018} and \cite{Lin2019} for the Riemannian functional PCA, we explicitly track the dependence of the convergence rate on the manifold's dimension $p$, absorbed in $c_{n}$ and $\epsilon_{n}$, instead of treating it as fixed. 
This distinction is crucial, as dimensionality often plays an important role in the convergence rates for the high-dimensional factor models.
In this regard, our result is reminiscent of the high-dimensional linear factor models for Euclidean time series, for which it was shown that $\kappa^{2} = O(p)$ and $\epsilon_{n} = O(p / \sqrt{n})$, assuming strong factors in the sense of  \cite{LYB2011}. 
Second, the manifold is not assumed to be complete.
In particular, this makes Theorem \ref{prop4} applicable to time series in the Bures--Wasserstein manifold.

The theoretical analysis requires bounding $\mathcal{P}_{\hat{\bm{\mu}}\rightarrow\bm{\mu}} \hat{\mathbf{z}}_{t} - \mathbf{z}_{t}$ non-asymptotically with careful geometric comparisons.
Standard tools used in the existing statistical analysis of manifold-valued data \citep{Bhattacharya2005, Bhattacharya2003, Wilfrid2011, Lin2019}, such as the covariant Taylor expansion of the log map, are inadequate here.
To overcome these challenges, we employ a holonomy argument that tracks changes in tangent vectors parallelly transported along small geodesic triangles. % (see the proof of Theorem \ref{prop3} in Appendix \ref{App::theory}).
To the best of our knowledge, this technique has not appeared in previous works.

Assumption \ref{ass:concentration} and Eq.~\eqref{P3-2} in Theorem \ref{prop4} are very weak assumptions on the serial dependence in the data, as they permit long-range dependent time series.
Theorem \ref{cor1} below refines Theorem \ref{prop4} and obtains a faster rate of convergence under an additional short-memory assumption.
Before presenting the result, we need to introduce some notations. 
For each $\bm{\xi} \in B_{\bm{\mu}}(R/(2+\eta))$, let $\mathbf{z}_{t,\bm{\xi}} = \rlog_{\bm{\xi}}(\mathbf{x}_{t})$.
The autocovariance operators of the process $\{\mathbf{z}_{t,\bm{\xi}}\}$ is denoted by 
\begin{align*}
    \mathbf{\Gamma}_{\bm{\xi}}(s, t) = \mathbb{E}\left[(\mathbf{z}_{s,\bm{\xi}} - \mathbb{E}\mathbf{z}_{s,\bm{\xi}}) \otimes (\mathbf{z}_{t,\bm{\xi}} - \mathbb{E}\mathbf{z}_{t,\bm{\xi}})\right].
\end{align*}
Now we can state Theorem \ref{cor1}.

\begin{theorem} \label{cor1}
Assume the same as in Theorem \ref{prop4}. Let $\mathbf{E} = (\mathbf{e}_{1}, \mathbf{e}_{2}, \ldots, \mathbf{e}_{p})$ be an orthonormal basis of $T_{\bm{\mu}}M$. Suppose each component of $\mathbf{z}_{t, \mathbf{E}}$, $z_{t,j} = \langle \mathbf{z}_{t}, \mathbf{e}_{j} \rangle$, $j = 1,2, \ldots, p$, admits an MA($\infty$) representation $z_{t,j} = \sum_{k=0}^{\infty} \alpha_{k,j} \epsilon_{t-k,j}$, where $\max_{1 \leq j \leq p} \sum_{k} \alpha_{k,j}^{2} = O(1)$ and $\{\epsilon_{t,j}\}$ is an i.i.d.~sequence of random variables such that
\begin{align*} 
    \sup_{\substack{j \in \{1,2,\ldots, p\} \\ -\infty < t < \infty}} \mathbb{E}(|\epsilon_{t,j}|^{4}) = O(1).
\end{align*}
In addition, there exists some $\epsilon > 0$ with the property that for each $\bm{\xi} \in B_{\bm{\mu}}(\epsilon)$, there exist $C_{\bm{\xi}} < \infty$ and $d_{\bm{\xi}} > 1$ such that
\begin{align} \label{cor1-statement-1}
    \frac{|\mathrm{Tr}(\mathbf{\Gamma}_{\bm{\xi}}(s,t))|}{\sqrt{\mathrm{Tr}(\mathbf{\Gamma}_{\bm{\xi}}(s,s)) \mathrm{Tr}(\mathbf{\Gamma}_{\bm{\xi}}(t,t)) }} \leq C_{\bm{\xi}} |s-t|^{-d_{\bm{\xi}}}
\end{align}
for all $s, t$, where
\begin{align} \label{cor1-statement-2}
    \sup_{\bm{\xi} \in B_{\bm{\mu}}(\epsilon)} C_{\bm{\xi}} = O(1), \mbox{ and} \quad  \inf_{n} \inf_{\bm{\xi} \in B_{\bm{\mu}}(\epsilon)} d_{\bm{\xi}} > 1. 
\end{align}
%and there exist $C_{j, \bm{\xi}}$ and $d_{j, \bm{\xi}}$ such that the autocovariance functions satisfy
%\begin{align} \label{cor1-statement-2}
%    |\gamma_{j,\bm{\xi}}(h)| \leq C_{j,\bm{\xi}} \gamma_{j,\bm{\xi}}(0) |h|^{-1 + 2 d_{j, \bm{\xi}}}, \quad j = 1, 2,\ldots, p,
%\end{align}
%where $d_{j, \bm{\xi}} \in (-0.5, 0)$ and 
%\begin{align} \label{cor1-statement-2.5}
%    \sup_{\substack{j=1,2,\ldots,p \\ \bm{\xi} \in B_{\bm{\mu}}(R/(2+\eta))}} C_{j, \bm{\xi}} = O(1), \quad  \sup_{n} \sup_{\substack{j=1,2,\ldots,p \\ \bm{\xi} \in B_{\bm{\mu}}(R/(2+\eta))}} d_{j, \bm{\xi}} < 0. 
%\end{align}
Then  
\begin{align}  \label{cor1-statement-3}
    \left\Vert \sin \Theta( \mathrm{span}\{\mathcal{P}_{\hat{\bm{\mu}} \rightarrow \bm{\mu}} \hat{\mathbf{a}}_{1} \ldots \mathcal{P}_{\hat{\bm{\mu}} \rightarrow \bm{\mu}} \hat{\mathbf{a}}_{r}\}, \mathrm{Im}(\mathcal{A}) ) \right\Vert_{op} = O_{p}\left(\frac{1}{\kappa^{2} \sqrt{n}}\right)
\end{align}
where $\kappa$ is as in Theorem~\ref{prop4} and $\sin\Theta(\cdot,\cdot)$ is defined after Eq.~\eqref{sinetheta_def}.
\end{theorem}

As shown in Theorem \ref{cor1}, under the short-memory condition \eqref{cor1-statement-1}--\eqref{cor1-statement-2} the estimated loading space converges at the dimension-free $n^{-1/2}$ rate if $\kappa \asymp 1$.
Thus, the Riemannian factor model can be estimated as efficiently as the high-dimensional linear factor model with strong factors \citep{LYB2011} and the fixed-dimensional Riemannian functional PCA \citep{Lin2019, dai2018}. 
We therefore refer to the case $\kappa \asymp 1$ as ``strong factors,'' and the case $\kappa = o(1)$ as ``weak factors'' for the Riemannian factor model.

It is worth pointing out that the dimension-free rate is not a straightforward extension from those earlier results.
In fact, both the signal strength and the estimation noise are quite different from the classical linear factor model.
Recall that in Theorem \ref{prop4}, the convergence rate for the estimated loading space is determined by the signal-to-noise ratio, in which the estimation noise is quantified by $\sqrt{c_{n}} + \epsilon_{n}$, and the signal strength by $\kappa^{2}$. 
For the signal strength, \ref{ass:support} implies
\begin{align*}
    \mathbb{E}\Vert \mathbf{z}_{t} \Vert_{\bm{\mu}}^{2} = \mathbb{E} \Vert \mathcal{A}\mathbf{f}_{t} \Vert_{\bm{\mu}}^{2} + \mathbb{E} \Vert \bm{\delta}_{t} \Vert_{\bm{\mu}}^{2} %\\
    = \mathbb{E} \Vert \mathbf{f}_{t} \Vert^{2} + \mathbb{E} \Vert \bm{\delta}_{t} \Vert_{\bm{\mu}}^{2} 
    \leq \left(\frac{R}{2+\eta}\right)^{2}.
\end{align*}
In particular, neither $\mathbb{E} \Vert \mathbf{f}_{t} \Vert^{2}$ nor $\mathbb{E} \Vert \bm{\delta}_{t} \Vert_{\bm{\mu}}^{2}$ can diverge with the dimension $p$, so the signal strength $\kappa^{2}$ is of order $O(1)$. 
For the noise component in Theorem \ref{prop4}, the term $c_{n}^{1/2}$ accounts for the estimation error of the Fr\'{e}chet mean $d_{M}(\hat{\bm{\mu}}, \bm{\mu})$, while the term $\epsilon_{n}$ reflects the estimation error of the autocovariance operators. 
Both errors, under the additional assumptions in Theorem \ref{cor1}, are of order $n^{-1/2}$, independent of $p$.
%Under the additional assumptions in Theorem \ref{cor1}, we show that both are of order $n^{-1/2}$, independent of $p$.

%{Reading through the proof, it seen that the additional assumption in 3.2 is mainly (or only) for $\mu$ and $c_n$. Should we separate the results for $\mu$ and the loading space? Because the new assumption in 3.2 is on covariance, I had the impression that it was for the loading space. But \eqref{P3-2} is always good as $p/n$ for finite maximum autocorrelation, and it cannot be reduced (am I right?). It can be rewritten in the loading space and the factor autocorrvariance. Only $\mu$ involves the sum of auto-correvariance, hence short/long memory is involved. Correct?}
The additional assumptions in Theorem \ref{cor1} are rather mild.
First, the i.i.d.~assumption on $\{\epsilon_{t,j}\}$ can be relaxed to allow for martingale difference sequences with conditional heteroscedasticity, at the expense of stronger moment conditions (see, e.g., Theorem A.2 in  \citealt{huang2025}).
Second, except when $\bm{\xi} = \bm{\mu}$, the process $\mathbf{z}_{t,\bm{\xi}}$ may not be weakly stationary.
Clearly, $\mathbf{z}_{t, \bm{\xi}}$ is stationary if $\{\mathbf{z}_{t}\}$ is strictly stationary, in which case $\mathbf{\Gamma}_{\bm{\xi}}(s,t)$ depends only on $|s-t|$.
Finally, condition \eqref{cor1-statement-1}, requiring the autocovariances of $\mathbf{z}_{t, \bm{\xi}}$ to decay sufficiently fast, ensures short-memory. 
It is natural in our setting because the autocovariance operator $\mathbf{\Gamma}_{\bm{\xi}}$ varies differentiably near $\bm{\mu}$. % and $\mathbf{z}_{t,\bm{\xi}}$ also varies differentiably near $\mathbf{z}_{t}$ \citep{docarmo1992, Lin2019}.
Moreover, it automatically holds when $\{\mathbf{z}_{t}\}$ is $m$-dependent for some finite $m$.
In that case, since $\mathbf{x}_{t}$ and $\mathbf{x}_{t-k}$ are independent for $k > m$, $\mathbf{\Gamma}_{\bm{\xi}}(s, t) = 0$ for all $|s - t| > m$. 

%as previously noted, condition \eqref{cor1-statement-1} is a short-memory condition which requires the autocovariances of $\mathbf{z}_{t, \bm{\xi}}$ to decay sufficiently fast. 
%Such condition is natural since 

Our final result, which gives convergence rates for the estimated eigenvalues, justifies the ratio estimator in Eq.~\eqref{Sec2-ratio} for selecting the number of factors. 
These rates also coincide with the results of \citet{LY2012} for the Euclidean time series. 

\begin{prop} \label{cor2}
    Under the assumptions of Theorem \ref{cor1}, we have
    \begin{itemize}
        \item[(i)] $|\hat{\lambda}_{i} - \lambda_{i}| = O_{p}(n^{-1/2})$ for $i = 1,2, \ldots, r$;
        \item[(ii)] $\hat{\lambda}_{i} = O_{p}(n^{-1})$ for $i > r$,
    \end{itemize}
    where $\lambda_{1} \geq \lambda_{2} \geq \ldots \geq \lambda_{p}$ are the eigenvalues of the operator $\mathbf{\Lambda} = \sum_{h=1}^{h_{0}} \mathbb{E}(\mathbf{z}_{t-h} \otimes \mathbf{z}_{t})\mathbb{E}(\mathbf{z}_{t-h} \otimes \mathbf{z}_{t})^{*}$. 
    In addition, $\lambda_{r} \geq \kappa^{2}$ and $\lambda_{i} = 0$ for $i > r$.
\end{prop}

\section{Simulation studies} \label{Sec::simulation}  \label{Sec::sim-pSpheres}
In this section, we assess the performance of the proposed RFM using synthetic data.
We consider data situated on two nonlinear manifolds, namely the Bures--Wasserstein manifold and the product manifold of spheres. 
Due to space constraints, we present the results for the Bures--Wasserstein manifold in Section {\bf S.3} of the supplementary material.
To illustrate the gains in capturing the underlying nonlinear geometry, we compare the RFM with the linear factor model (LFM) in terms of the pseudo-prediction errors. 
The results show that the RFM can achieve a low pseudo-prediction error with a few factors.
In addition, when the sample size is sufficiently large, the selected number of factors and the estimated loading space are also accurate, supporting the theoretical results.

In the following, we consider sphere-valued time series.
Specifically, we apply the proposed RFM to the more complex setup where the time series take values in the product of unit spheres $M = \mathbb{S}^{q_{1}} \times \ldots \times \mathbb{S}^{q_{d}}$, equipped with the product metric, where $\mathbb{S}^{q}$ denotes the unit sphere in $\mathbb{R}^{q+1}$.
Sphere-valued data are commonly seen in environmental studies, such as wind directions or ocean currents, as well as in statistical shape analysis \citep{dryden2016}. They have also attracted much attention due to its connection to compositional data \citep{WANG2007, dai2018}. 

The synthetic data are generated as follows. 
The reference point $\bm{\mu} = (\bm{\mu}_{1}, \ldots, \bm{\mu}_{d}) \in M$ has each of its components $\bm{\mu}_{j} \in \mathbb{S}^{q_{j}}$ uniformly selected on its corresponding sphere. 
Let $\mathbf{f}_{t} \in \mathbb{R}^{r}$ be the factor process, which follows an AR(1) model
\begin{align} \label{sim--1}
    \mathbf{f}_{t} = \alpha \mathbf{f}_{t-1} + s \cdot \bm{\epsilon}_{t},
\end{align}
where $|\alpha|<1$, $s > 0$ and $\bm{\epsilon}_{t} = (\epsilon_{t,1}, \ldots, \epsilon_{t,r})^{\top}$ is drawn independently from the standard normal distribution truncated at the unit ball to ensure bounded support.
In this subsection, we set $s = 2.5\sqrt{d / 5}$ and $\alpha = 0.5$.
Let $\mathbf{E} = (\mathbf{e}_{1}, \ldots, \mathbf{e}_{p})$ be an orthonormal basis of $T_{\bm{\mu}}M$, and $\mathbf{A}_{\mathbf{E}} \in \mathbb{R}^{p \times r}$, where $p = q_{1} + \ldots + q_{d}$ is the manifold dimension, be a matrix with randomly generated orthonormal columns.
In addition, $\bm{\delta}_{t, \mathbf{E}}$ is independently generated from the uniform distribution on the $p$-sphere of radius $\sigma_{z}$ and set $\mathbf{z}_{t, \mathbf{E}} = \mathbf{A}_{\mathbf{E}} \mathbf{f}_{t} + \bm{\delta}_{t, \mathbf{E}}$.
Finally, the observed data $\mathbf{x}_{t}$ is obtained through the exponential map $\mathbf{x}_{t} = \rexp_{\bm{\mu}}(\mathbf{z}_{t})$, where $\mathbf{z}_{t} = J_{\mathbf{E}}(\mathbf{z}_{t,\mathbf{E}})$ with $J_{\mathbf{E}}$ defined in \eqref{J_def}.
We consider the cases $d \in \{5, 10, 20\}$ and set $q_{j} = 4$ for $j = 1,2, \ldots, d$.

To illustrate the impact of the nonlinear constraint of the manifold, we compare the RFM with the linear factor model (LFM) of \citet{LYB2011}. In practice, it is not uncommon to treat $\mathbf{x}_{t,j}$ as Euclidean data in the usual coordinates of $\mathbb{R}^{q_{j}+1}$, denoted hereafter by $\mathcal{E}(\mathbf{x}_{t,j}) \in \mathbb{R}^{q_{j}+1}$.
Hence, in the following experiments, we apply the LFM to $\mathcal{E}(\mathbf{x}_{t}) = (\mathcal{E}(\mathbf{x}_{t,1}), \ldots, \mathcal{E}(\mathbf{x}_{t,d})) \in \mathbb{R}^{\sum_{j=1}^{d}(q_{j}+1)}$. 

For each experiment, a test set of size $n_{\mathrm{test}} = 200$ is reserved to evaluate the performance.
Let $\hat{\mathbf{x}}^{(L)}_{t}$ and $\hat{\mathbf{x}}^{(R)}_{t}$ be the pseudo-predictions of LFM and RFM, respectively.
Specifically, at time $t$, given a new observation $\mathbf{x}_{t}$, the pseudo-prediction of RFM is given by $\hat{\mathbf{x}}_{t}^{(R)} = \rexp_{\hat{\bm{\mu}}}(J_{\hat{\mathbf{E}}}(\hat{\mathbf{A}}_{\hat{\mathbf{E}}}\hat{\mathbf{A}}_{\hat{\mathbf{E}}}^{\top}\hat{\mathbf{z}}_{t, \hat{\mathbf{E}}}))$, where $\hat{\bm{\mu}}$ is the estimated Fr\'{e}chet mean, $\hat{\mathbf{E}}$ is an orthonormal basis of $T_{\hat{\bm{\mu}}}M$, $\hat{\mathbf{z}}_{t, \hat{\mathbf{E}}} = J_{\hat{\mathbf{E}}}^{-1}(\rlog_{\hat{\bm{\mu}}}(\mathbf{x}_{t}))$, and $\hat{\mathbf{A}}_{\hat{\mathbf{E}}}$ is the estimated loading matrix using the training sample $\mathbf{x}_1, \ldots, \mathbf{x}_{n}$. 
Similarly, the pseudo-predictions for LFM are computed as $\mathcal{E}(\tilde{\mathbf{x}}_{t}^{(L)}) = \tilde{\bm{\mu}} + \tilde{\mathbf{A}} \tilde{\mathbf{A}}^{\top} \tilde{\mathbf{z}}_{t}$, where $\tilde{\mathbf{z}}_{t} = \mathcal{E}(\mathbf{x}_{t}) - \tilde{\bm{\mu}}$, $\tilde{\bm{\mu}}$ is the Euclidean mean of $\mathcal{E}(\mathbf{x}_{1}), \ldots, \mathcal{E}(\mathbf{x}_{n})$, and $\tilde{\mathbf{A}} \in \mathbb{R}^{p \times k}$ is the loading matrix estimated by LFM with $k$ being the number of extracted factors. 
Note that $\mathcal{E}(\tilde{\mathbf{x}}_{t}^{(L)})$, the predictions of LFM, is not on the sphere in general.
Thus, $\mathcal{E}(\hat{\mathbf{x}}_{t}^{(L)})$ is obtained by projecting $\mathcal{E}(\tilde{\mathbf{x}}_{t}^{(L)})$ to the unit spheres. 
Then, the relative geodesic mean squared error (rGMSE), defined as
\begin{align} \label{sim-0}
    \mbox{rGMSE} = \frac{\sum_{t=n + 1}^{n + n_{\mathrm{test}}} d_{M}(\mathbf{x}_{t}, \hat{\mathbf{x}}_{t})^{2}}{\sum_{t=n+1}^{n + n_{\mathrm{test}}} d_{M}(\mathbf{x}_{t}, \hat{\bm{\mu}})^{2}},
\end{align}
is used to evaluate the methods.

As shown in Table \ref{Sec4-tab-3}, LFM and RFM achieve similar rGMSEs when fewer than 5 factors are used.
However, when $r = 5$ factors are employed, RFM outperforms LFM, reducing the rGMSE by at least 20\% compared to the LFM. 
In addition, to achieve the same level of rGMSE as the RFM with $r$ factors, the LFM requires four or more additional factors.
This highlights the effect of the manifold's nonlinearity on LFM, which cannot be easily compensated by simply increasing the number of factors.
In addition, Table \ref{sec-4-tab4} shows the frequency with which the ratio estimator correctly identifies the number of factors.
The correct number of factors is selected with high frequency---exceeding 80\% of the time when $n = 100$ and near 100\% when $n=200$ in all cases, validating the efficacy of the ratio estimator in selecting the number of factors for the RFM.

\spacingset{1}
\begin{table}[t]
\centering
\caption{Relative GMSE of LFM and RFM applied to the product of spheres. The results are averaged over 300 simulations. The true number of factors $r = 5$.}
\label{Sec4-tab-3}
\begin{tabular}{@{}lrrrrrrrrrr@{}} 
\toprule 
Num.~of factors & 1 & 2 & 3 & 4 & 5 & 6 & 7 & 8 & 9 & 10 \\ \midrule
\multicolumn{11}{l}{$(n, d) = (50, 5)$} \\
LFM & 0.78 & 0.61 & 0.46 & 0.33 & 0.23 & 0.19 & 0.16 & 0.14 & 0.13 & 0.12 \\
RFM & 0.78 & 0.60 & 0.43 & 0.28 & 0.15 & 0.13 & 0.12 & 0.11 & 0.10 & 0.09 \\
\multicolumn{11}{l}{$(n, d) = (100, 5)$} \\
LFM & 0.78 & 0.60 & 0.44 & 0.31 & 0.21 & 0.17 & 0.14 & 0.13 & 0.11 & 0.10 \\
RFM & 0.78 & 0.59 & 0.42 & 0.27 & 0.13 & 0.12 & 0.11 & 0.10 & 0.09 & 0.08 \\ 
\multicolumn{11}{l}{$(n, d) = (200, 5)$} \\
LFM & 0.77 & 0.59 & 0.44 & 0.30 & 0.19 & 0.16 & 0.14 & 0.12 & 0.11 & 0.10 \\
RFM & 0.78 & 0.58 & 0.41 & 0.26 & 0.12 & 0.11 & 0.10 & 0.09 & 0.08 & 0.07 \\\hline
\multicolumn{11}{l}{$(n, d) = (50, 10)$} \\
LFM & 0.79 & 0.62 & 0.47 & 0.35 & 0.25 & 0.22 & 0.20 & 0.19 & 0.18 & 0.17 \\ 
RFM & 0.79 & 0.61 & 0.45 & 0.31 & 0.18 & 0.17 & 0.16 & 0.15 & 0.15 & 0.14 \\
\multicolumn{11}{l}{$(n, d) = (100, 10)$} \\
LFM & 0.78 & 0.61 & 0.46 & 0.33 & 0.23 & 0.19 & 0.18 & 0.17 & 0.16 & 0.15 \\ 
RFM & 0.79 & 0.61 & 0.44 & 0.29 & 0.16 & 0.15 & 0.14 & 0.13 & 0.13 & 0.12 \\
\multicolumn{11}{l}{$(n, d) = (200, 10)$} \\
LFM & 0.78 & 0.60 & 0.45 & 0.32 & 0.21 & 0.18 & 0.17 & 0.16 & 0.15 & 0.14 \\ 
RFM & 0.79 & 0.60 & 0.43 & 0.28 & 0.14 & 0.14 & 0.13 & 0.13 & 0.12 & 0.12 \\ \hline
\multicolumn{11}{l}{$(n, d) = (50, 20)$} \\
LFM & 0.79 & 0.62 & 0.48 & 0.36 & 0.26 & 0.23 & 0.22 & 0.21 & 0.20 & 0.19 \\ 
RFM & 0.80 & 0.62 & 0.47 & 0.32 & 0.20 & 0.19 & 0.18 & 0.18 & 0.17 & 0.17 \\ 
\multicolumn{11}{l}{$(n, d) = (100, 20)$} \\
LFM & 0.79 & 0.61 & 0.47 & 0.34 & 0.24 & 0.21 & 0.20 & 0.19 & 0.18 & 0.18 \\ 
RFM & 0.80 & 0.61 & 0.45 & 0.30 & 0.17 & 0.16 & 0.16 & 0.16 & 0.15 & 0.15 \\ 
\multicolumn{11}{l}{$(n, d) = (200, 20)$} \\
LFM & 0.79 & 0.61 & 0.46 & 0.33 & 0.22 & 0.20 & 0.19 & 0.18 & 0.17 & 0.17 \\ 
RFM & 0.80 & 0.61 & 0.44 & 0.29 & 0.16 & 0.15 & 0.15 & 0.15 & 0.14 & 0.14 \\ \bottomrule
\end{tabular}
\end{table}
\spacingset{1.8}

\spacingset{1}
\begin{table}[th]
\centering
\caption{Proportion of times the correct number of factors is selected over 300 simulations. The data are on the product sphere $M = \mathbb{S}^{q_{1}} \times \ldots \times \mathbb{S}^{q_{d}}$ generated according to Section \ref{Sec::sim-pSpheres}.}
\label{sec-4-tab4}
\begin{tabular}{cccc}
\toprule
\diagbox[width=3.5em, height=1.5em]{$n$}{$d$} & 5 & 10 & 20 \\ 
\midrule
50  & 0.57 & 0.61 & 0.60   \\
100 & 0.83 & 0.88 & 0.91   \\
200 & 0.99 & 0.98 & 0.99   \\ \bottomrule
\end{tabular}
\end{table}
\spacingset{1.8}

\section{Empirical application} \label{Sec::application}
In this section, we apply the proposed RFM to U.S.~financial data. 
In particular, we view the time-varying covariances (realized volatility) of the daily log returns of a selected set of U.S.~stocks as a time series in the Bures--Wasserstein manifold. 
In addition to modeling the time series,
we also investigate the out-of-sample forecasting performance of the RFM, which has important practical implications. 
The data used in the analysis were obtained from Yahoo!~Finance, and the codes for data acquisition and analysis are available in the Github repository [Omitted for anonymity].%(\url{https://github.com/shuochieh/Riemannian_factor_model}).

We consider the daily log returns of the S\&P 500 component stocks during January 1, 2000 to December 31, 2019, and retain only the companies with complete trading records in this period.
Among them, we focus on the four largest companies by market capitalization from each of the three sectors: information technology (IT), finance, and energy.
To keep the selection more homogeneous, we exclude Nvidia and Berkshire Hathaway, whose market dynamics differ from their corresponding sectors in ways that would complicate comparisons.
%{Should we mention we removed NVDA and BRK.B?}
The 12 selected companies, along with their tickers we used throughout the section, are listed in Table \ref{AppC-tab1}.
For each month, we compute the realized covariance matrix $\mathbf{\Sigma}_{t} \in \mathbb{R}^{12 \times 12}$ of the daily log returns (in percentage points), yielding 240 monthly matrix observations. The first 204 months are used as the training set and the final 36 months are reserved as test data.

\spacingset{1}
\begin{table}[t]
\centering
\caption{The companies and their tickers used in the real data analysis.}
\label{AppC-tab1}
\begin{tabular}{@{}llllll@{}}
\toprule
Company & Ticker & Company & Ticker & Company & Ticker \\ \midrule
Microsoft          & MSFT  &  JPMorgan Chase     & JPM    & ExxonMobil          & XOM  \\
Apple Inc.         & AAPL  &  Bank of America    & BAC    & Chevron Corporation & CVX  \\
Oracle Corporation & ORCL  &  Wells Fargo        & WFC    & ConocoPhillips      & COP  \\
Cisco              & CSCO  &  Goldman Sachs      & GS     & EOG Resources       & EOG   \\ \bottomrule
\end{tabular}
\end{table}
\spacingset{1.8}

\begin{figure}[th] 
    \centering
    \includegraphics[width=\textwidth]{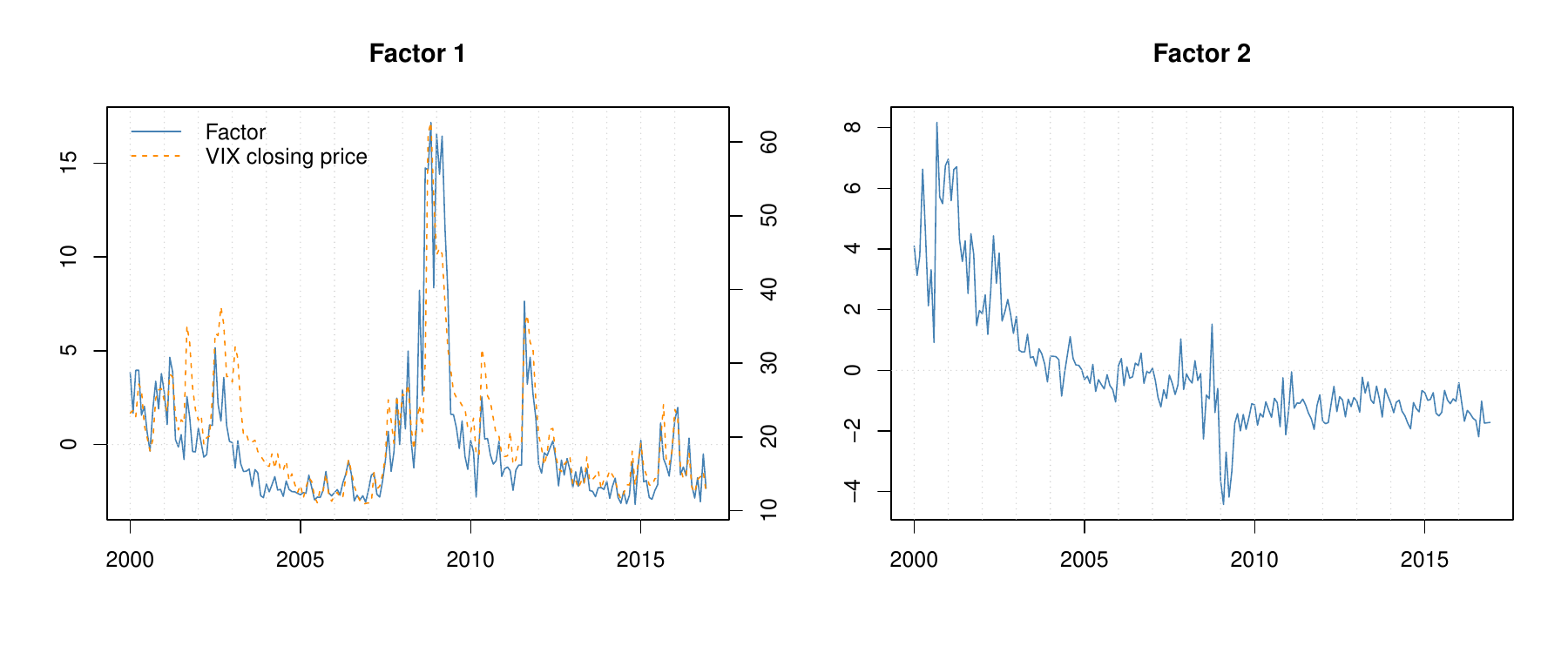}
    \caption{The leading two estimated factors (blue lines) from the U.S.~financial data. The orange dashed line indicates the monthly average of VIX closing price.} 
    \label{sec-5-fig1}
\end{figure}

\begin{figure}[ht]
    \centering
    \includegraphics[width=\textwidth]{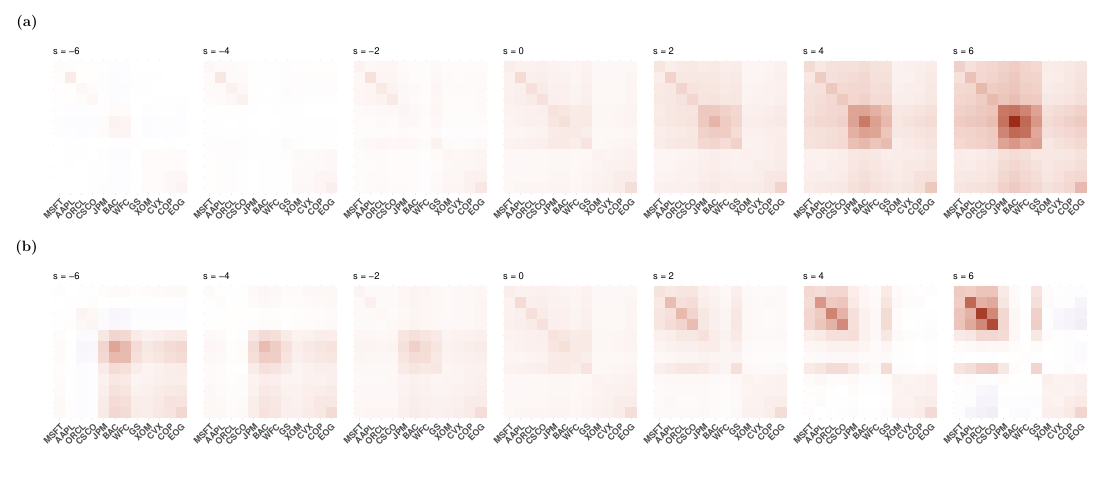}
    \caption{The loadings corresponding to the first factor (row (a)) and the second factor (row (b)) estimated from the U.S.~return data, represented as the geodesic trajectory $\rexp_{\hat{\bm\Sigma}}(s\hat{\mathbf{a}}_j)$ for $s\in[-6,6]$, $j = 1, 2$, on the Bures--Wasserstein manifold. Red and blue entries signify positive and negative values, with darker colors representing larger magnitude. The color white corresponds to zero.}
    \label{sec-5-fig2}
\end{figure}

First, we estimate the RFM and LFM using the training set and construct the pseudo-predictions as in Section \ref{Sec::sim-pSpheres} (for LFM applied to SPD matrices, see Section S.3 of the supplementary material).
Table~\ref{sec-5-tab1} reports the relative geodesic mean squared errors (rGMSE) defined in \eqref{sim-0}. 
The LFM attains an rGMSE of 0.405 using 9 factors. 
In contrast, the geometry-aware RFM achieves an rGMSE of 0.404 with only two factors.
Figure \ref{sec-5-fig1} plots the leading factors estimated by the RFM. 
The first leading factor shows a pronounced spike for the 2008 financial crisis. The second factor has relatively larger values during the dot-com bubble, and it moves in the opposite direction during the financial crisis.
To visualize and interpret the effects of these factors in the manifold, one can examine the geodesic trajectories traveled in the direction of the corresponding loading vectors. Specifically, given the estimated factor loading $\hat{\mathbf{a}}_{j}$ in the tangent space, the geodesic trajectory $\hat{\mathbf{\Sigma}}_{(j)}(s) = \rexp_{\hat{\mathbf{\Sigma}}}(s \hat{\mathbf{a}}_{j})$, $s \in \mathbb{R}$, where $\hat{\mathbf{\Sigma}}$ is the estimated Fr\'{e}chet mean, shows the effect of the $j$-th factor on the prediction, holding all other factors at zero. 
Figure \ref{sec-5-fig2} presents the resulting trajectories as heatmaps of the entries of $\hat{\mathbf{\Sigma}}_{(j)}(s)$.
The first factor primarily reflects the intensity of the interdependence between the selected companies, especially the financial companies.
During crises, return correlations tend to be more elevated, and this factor captures such broad market-wide effect.
Indeed, as shown in Figure \ref{sec-5-fig1}, the first factor closely tracks the closing price of the VIX volatility index, a widely-used proxy for aggregate market uncertainty.
Meanwhile, the second factor captures sector-specific volatility dynamics. 
As shown in Figure \ref{sec-5-fig2}(b), positive values of the second factor, such as during the dot-com bubble, indicate heightened dependence within the IT sector.
Negative values, such as during the financial crisis, shift the loading concentration to the financial and energy sectors. 

\spacingset{1}
\begin{table}[t]
\centering
\caption{Relative geodesic mean squared errors (rGMSE) of LFM and RFM applied to U.S.~return data.}
\label{sec-5-tab1} 
\begin{tabular}{@{}lrrrrrrrrrr@{}}
\toprule
Num.~of factors & 1 & 2 & 3 & 4 & 5 & 6 & 7 & 8 & 9 & 10 \\ \midrule
LFM & 1.183 & 0.533 & 0.455 & 0.429 & 0.423 & 0.410 & 0.406 & 0.410 & 0.405 & 0.391 \\
RFM & 0.539 & 0.404 & 0.344 & 0.323 & 0.295 & 0.267 & 0.258 & 0.244 & 0.238 & 0.229 \\ \bottomrule
\end{tabular}
\end{table}
\spacingset{1.8}

\spacingset{1}
\spacingset{1.8}

\spacingset{1}
\spacingset{1.8}

\begin{figure}[t] 
    \centering
    \includegraphics[width=\textwidth]{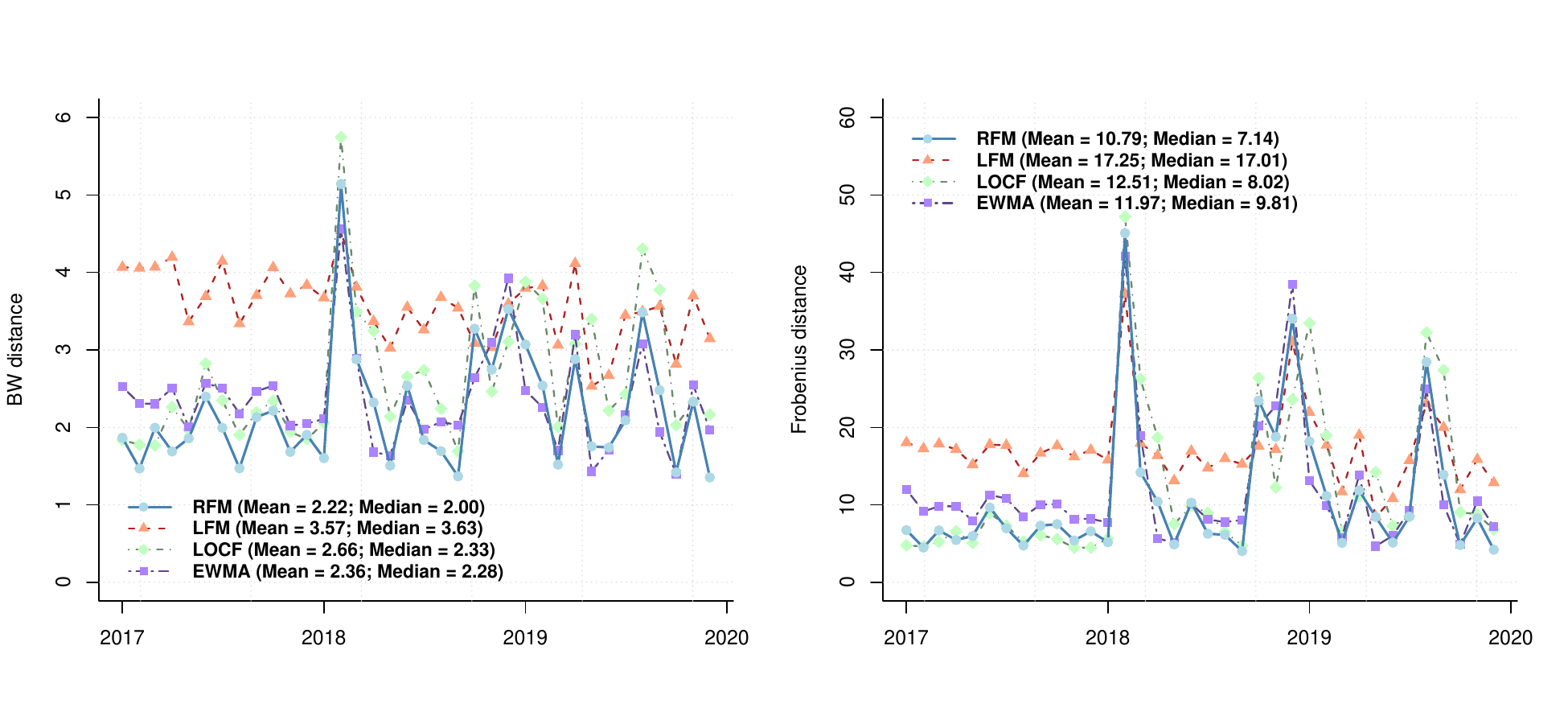}
    \caption{Out-of-sample prediction errors of RFM, LFM, LOCF, and EWMA. Left: In Bures--Wasserstein distance. Right: In Frobenius distance.}
    \label{sec-5-fig3}
\end{figure}

Next, we evaluate the out-of-sample forecasting performance.
We fit a VAR(1) model to the estimated factors using data up until time $t$, and obtain the one-step-ahead forecast $\hat{\mathbf{f}}_{t+1}$ and $\hat{\mathbf{\Sigma}}_{t+1}$. 
The number of factors used in forecasting is selected by the ratio estimator in Eq.~\eqref{Sec2-ratio}.
As additional benchmarks, we employ two simple and widely used methods.
The first is the last observation carried forward (LOCF) method, which uses the previous month's covariance matrix as the forecast.
The second is the exponentially weighted moving average (EWMA) method, implemented in RiskMetrics (\citealp{longerstaey1996}; see also \citealp{Kasper2023}) developed by J.~P.~Morgan.
Specifically, we implement the EWMA by the recursive formula $\hat{\mathbf{\Sigma}}_{t+1} = \lambda \hat{\mathbf{\Sigma}}_{t} + (1 - \lambda) \mathbf{\Sigma}_{t}$, where $\lambda$ is the forgetting factor which is set to 0.94 by the rule of thumb, and $\hat{\mathbf{\Sigma}}_{1} = \mathbf{0}$.
Figure \ref{sec-5-fig3} shows the out-of-sample prediction errors in the Bures--Wasserstein and in the Frobenius distance.
In both cases, RFM consistently yields competitive predictions, whereas LFM suffers from the largest errors. 
Both the LOCF and EWMA are slightly inferior to the RFM, indicating the proposed method is a viable alternative to these popular benchmarks. 

\begin{figure}[th] 
    \centering
    \includegraphics[width=\textwidth]{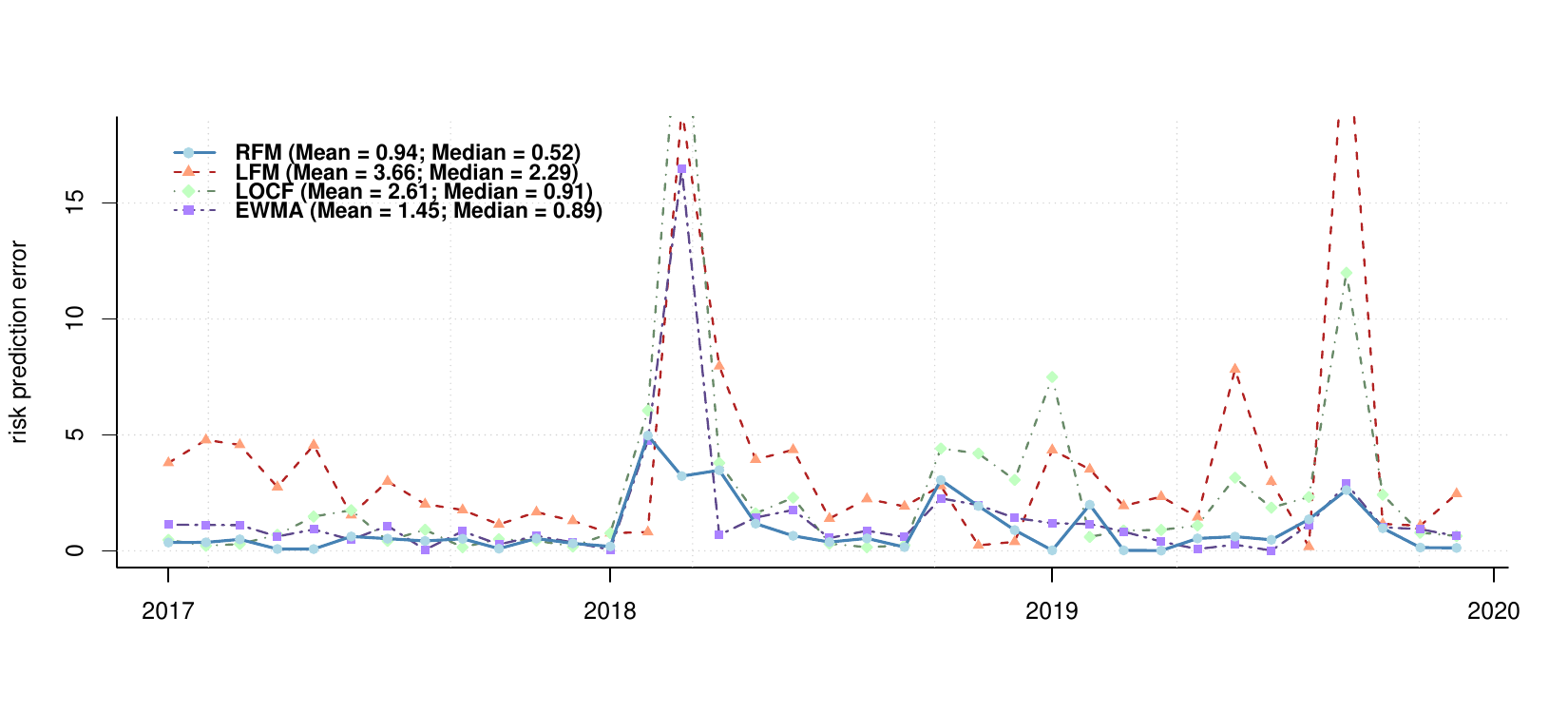}
    \caption{Out-of-sample risk prediction errors of RFM, LFM, LOCF, and EWMA for the minimum variance portfolios.}
    \label{sec-5-fig4}
\end{figure}

Since covariance forecasting is often used in portfolio risk assessments, we also evaluate the methods in this context.
For illustration, we consider the minimum variance portfolio $\mathbf{w}_{t}$ at each time $t$, which is given by $\mathbf{w}_{t} = \mathbf{\Sigma}_{t}^{-1}\mathbf{1}/(\mathbf{1}^{\top}\mathbf{\Sigma}_{t}^{-1}\mathbf{1})$ if short positions are allowed. 
The predicted risk of this portfolio for the following month is given by $\mathbf{w}_{t}^{\top}\hat{\mathbf{\Sigma}}_{t+1}\mathbf{w}_{t}$.
Figure \ref{sec-5-fig4} shows the risk prediction errors $|\mathbf{w}_{t}^{\top}(\hat{\mathbf{\Sigma}}_{t+1} - \mathbf{\Sigma}_{t+1}) \mathbf{w}_{t}|$ across the test period.
The RFM consistently yields low prediction errors throughout, with a mean error of 0.94, substantially outperforming LFM (3.66), LOCF (2.61), and EWMA (1.45).
The predictions of RFM are also quite robust, achieving a median error of 0.52, the lowest among all methods, compared to the second-best EWMA (0.89).
%The mean error of EWMA (1.28) is only slightly higher than the RFM, but its median error (0.96) is notably larger than that of RFM (0.48), the lowest among all methods.
These results demonstrate the effectiveness of RFM in risk assessments.
To summarize, in the analysis of the realized volatility series, the RFM not only offers effective dimension reduction with interpretable factors, but also produces competitive out-of-sample forecasts. This, in turn, facilitates accurate risk assessments, corroborating the proposed method's practical value. 

\spacingset{1}
\spacingset{1.8}

\spacingset{1}
\spacingset{1.8}

\section{Conclusion} \label{Sec::conclusion}

In this paper, we proposed the Riemannian factor model, a new factor modeling approach for time series in nonlinear manifolds.
We established the convergence rates for the estimated loading space, which, under short-memory and strong factor conditions, can be as fast as the dimension-free  rate $n^{-1/2}$, matching the ideal convergence rates for the conventional linear factor models.
Numerical experiments, including simulation studies and an analysis with the realized covariance matrices of the U.S.~stock returns, show that the RFM offers efficient dimension reduction, interpretable factors, and enhanced prediction capacity, which supports the RFM as a viable factor modeling approach for manifold-valued time series.

%\section{Disclosure statement}\label{disclosure-statement}

%The authors report there are no competing interests to declare.

\section{Data Availability Statement}\label{data-availability-statement}

The data used in the analysis were obtained from Yahoo!~Finance, which cannot be released according to its API's terms of use. However, the codes for data acquisition and analysis are available in the Github repository \url{https://github.com/shuochieh/Riemannian_factor_model}.

\spacingset{1}
\bibliography{reference}

@article{dai2018,
    author = {Xiongtao Dai and Hans-Georg M{\"u}ller},
    title = {Principal component analysis for functional data on {R}iemannian manifolds and spheres},
    volume = {46},
    journal = {The Annals of Statistics},
    number = {6B},
    pages = {3334--3361},
    year = {2018}
}

@article{Bhattacharya2003,
    author = {Rabi Bhattacharya and Vic Patrangenaru},
    title = {Large sample theory of intrinsic and extrinsic sample means on manifolds. {I}},
    volume = {31},
    journal = {The Annals of Statistics},
    number = {1},
    pages = {1--29},
    year = {2003}
}

@article{Bhattacharya2005,
    author = {Rabi Bhattacharya and Vic Patrangenaru},
    title = {Large sample theory of intrinsic and extrinsic sample means on manifolds--{II}},
    volume = {33},
    journal = {The Annals of Statistics},
    number = {3},
    pages = {1225--1259},
    year = {2005}
}

@book{docarmo1992,
  author       = {Manfredo P. do Carmo},
  title        = {Riemannian Geometry},
  series       = {Mathematics: Theory \& Applications},
  publisher    = {Birkhäuser Boston},
  year         = {1992}
}

@article{LYB2011,
 author = {Clifford Lam and Qiwei Yao and Neil Bathia},
 journal = {Biometrika},
 number = {4},
 pages = {901--918},
 title = {Estimation of latent factors for high-dimensional time series},
 volume = {98},
 year = {2011}
}

@article{LY2012,
    author = {Clifford Lam and Qiwei Yao},
    title = {Factor modeling for high-dimensional time series: Inference for the number of factors},
    volume = {40},
    journal = {The Annals of Statistics},
    number = {2},
    pages = {694--726},
    year = {2012}
}

@article{Lin2019,
    author = {Zhenhua Lin and Fang Yao},
    title = {Intrinsic {R}iemannian functional data analysis},
    volume = {47},
    journal = {The Annals of Statistics},
    number = {6},
    pages = {3533--3577},
    year = {2019}
}

@article{davis1970,
    author = {Davis, Chandler and Kahan, W. M.},
    title = {The Rotation of Eigenvectors by a Perturbation. {III}},
    journal = {SIAM Journal on Numerical Analysis},
    volume = {7},
    number = {1},
    pages = {1--46},
    year = {1970}
}

@article{chen2021,
    year = {2021},
    volume = {14},
    journal = {Foundations and Trends{\textregistered} in Machine Learning},
    title = {Spectral Methods for Data Science: A Statistical Perspective},
    number = {5},
    pages = {566--806},
    author = {Yuxin Chen and Yuejie Chi and Jianqing Fan and Cong Ma}
}

@article{Malago2018,
  author    = {Luigi Malag\`o and Luigi Montrucchio and Giovanni Pistone},
  title     = {Wasserstein {R}iemannian geometry of {G}aussian densities},
  journal   = {Information Geometry},
  year      = {2018},
  volume    = {1},
  number    = {2},
  pages     = {137--179}
}

@article{Takatsu2011,
    author = {Asuka Takatsu},
    title = {{Wasserstein geometry of Gaussian measures}},
    volume = {48},
    journal = {Osaka Journal of Mathematics},
    number = {4},
    pages = {1005--1026},
    year = {2011}
}

@book{Chewi2025,
  year = {2025},
  publisher = {Springer Cham},
  series = {Lecture Notes in Mathematics},
  title = {Statistical Optimal Transport},
  author = {Sinho Chewi and Jonathan Niles-Weed and Philippe Rigollet},
  edition = {1st} 
}

@article{Moakher2005,
    author = {Moakher, Maher},
    title = {A Differential Geometric Approach to the Geometric Mean of Symmetric Positive-Definite Matrices},
    journal = {SIAM Journal on Matrix Analysis and Applications},
    volume = {26},
    number = {3},
    pages = {735--747},
    year = {2005}
}

@article{Pennec2006,
  author    = {Xavier Pennec and Pierre Fillard and Nicholas Ayache},
  title     = {A {R}iemannian Framework for Tensor Computing},
  journal   = {International Journal of Computer Vision},
  year      = {2006},
  volume    = {66},
  number    = {1},
  pages     = {41--66}
}

@article{Arsigny2007,
    author = {Arsigny, Vincent and  Fillard, Pierre and  Pennec, Xavier and  Ayache, Nicholas},
    title = {Geometric Means in a Novel Vector Space Structure on Symmetric Positive‐Definite Matrices},
    journal = {SIAM Journal on Matrix Analysis and Applications},
    volume = {29},
    number = {1},
    pages = {328--347},
    year = {2007}
}

@ARTICLE{Fletcher2004,
  author={Fletcher, P.T. and Conglin Lu and Pizer, S.M. and Sarang Joshi},
  journal={IEEE Transactions on Medical Imaging}, 
  title={Principal geodesic analysis for the study of nonlinear statistics of shape}, 
  year={2004},
  volume={23},
  number={8},
  pages={995--1005}
}

@article{Fletcher2007,
    title = {Riemannian geometry for the statistical analysis of diffusion tensor data},
    journal = {Signal Processing},
    volume = {87},
    number = {2},
    pages = {250--262},
    year = {2007},
    author = {P. Thomas Fletcher and Sarang Joshi}
}

@article{Huckemann2010,
  author    = {Stephan Huckemann and Thomas Hotz and Axel Munk},
  title     = {Intrinsic Shape Analysis: Geodesic {PCA} for {R}iemannian Manifolds Modulo Isometric {L}ie Group Actions},
  journal   = {Statistica Sinica},
  volume    = {20},
  number    = {1},
  pages     = {1--100},
  year      = {2010}
}

@ARTICLE{Harandi2018,
  author={Harandi, Mehrtash and Salzmann, Mathieu and Hartley, Richard},
  journal={IEEE Transactions on Pattern Analysis and Machine Intelligence}, 
  title={Dimensionality Reduction on SPD Manifolds: The Emergence of Geometry-Aware Methods}, 
  year={2018},
  volume={40},
  number={1},
  pages={48--62}
}

@book{Wainwright_2019, 
    series={Cambridge Series in Statistical and Probabilistic Mathematics}, 
    title={High-Dimensional Statistics: A Non-Asymptotic Viewpoint}, 
    publisher={Cambridge University Press}, 
    author={Wainwright, Martin J.}, 
    year={2019}, 
    collection={Cambridge Series in Statistical and Probabilistic Mathematics}
}

@book{Buhlmann2011,
  author    = {Peter B{\"u}hlmann and Sara van de Geer},
  title     = {Statistics for High-Dimensional Data: Methods, Theory and Applications},
  series    = {Springer Series in Statistics},
  publisher = {Springer},
  year      = {2011}
}

@book{Boucheron,
    author = {Boucheron, Stéphane and Lugosi, Gábor and Massart, Pascal},
    title = {Concentration Inequalities: A Nonasymptotic Theory of Independence},
    publisher = {Oxford University Press},
    year = {2013}
}

@article{Chen2018,
  author  = {Likai Chen and Wei Biao Wu},
  title   = {Concentration inequalities for empirical processes of linear time series},
  journal = {Journal of Machine Learning Research},
  year    = {2018},
  volume  = {18},
  number  = {231},
  pages   = {1--46}
}

@misc{deb2024,
      title={Trade-off Between Dependence and Complexity for Nonparametric Learning -- an Empirical Process Approach}, 
      author={Nabarun Deb and Debarghya Mukherjee},
      year={2024},
      note={arXiv preprint arXiv:2401.08978},
      url={https://arxiv.org/abs/2401.08978} 
}

@article{huang2025,
  title={Model Selection for Unit-root Time Series with Many Predictors},
  author={Huang, Shuo-Chieh and Ing, Ching-Kang and Tsay, Ruey S},
  journal={arXiv preprint arXiv:2505.04884},
  year={2025}
}

@article{longerstaey1996,
  title={Risk Metrics},
  author={Longerstaey, Jacques and Zangari, Peter},
  year={1996}
}

@article{Kasper2023,
    year = {2023},
    volume = {12},
    journal = {Foundations and Trends{\textregistered} in Econometrics},
    title = {A Simple Method for Predicting Covariance Matrices of Financial Returns},
    doi = {10.1561/0800000047},
    number = {4},
    pages = {324--407},
    author = {Kasper Johansson and Mehmet G. Ogut and Markus Pelger and Thomas Schmelzer and Stephen Boyd}
}

@article{Forni2000,
    author = {Forni, Mario and Hallin, Marc and Lippi, Marco and Reichlin, Lucrezia},
    title = {The Generalized Dynamic-Factor Model: Identification and Estimation},
    journal = {The Review of Economics and Statistics},
    volume = {82},
    number = {4},
    pages = {540--554},
    year = {2000},
    month = {11},
    doi = {10.1162/003465300559037}
}

@incollection{Stock2011,
    author = {Stock, James H. and Watson, Mark W.},
    title = {Dynamic Factor Models},
    booktitle = {The Oxford Handbook of Economic Forecasting},
    publisher = {Oxford University Press},
    year = {2011},
    doi = {10.1093/oxfordhb/9780195398649.013.0003}
}

@article{bai2008,
  title={Large dimensional factor analysis},
  author={Bai, Jushan and Ng, Serena},
  journal={Foundations and Trends{\textregistered} in Econometrics},
  volume={3},
  number={2},
  pages={89--163},
  year={2008}
}

@incollection{Stock2006,
    title = {Forecasting with Many Predictors},
    booktitle = {Handbook of Economic Forecasting},
    publisher = {Elsevier},
    volume = {1},
    pages = {515--554},
    year = {2006},
    doi = {https://doi.org/10.1016/S1574-0706(05)01010-4},
    author = {James H. Stock and Mark W. Watson}
}

@article{bai2002,
  title={Determining the number of factors in approximate factor models},
  author={Bai, Jushan and Ng, Serena},
  journal={Econometrica},
  volume={70},
  number={1},
  pages={191--221},
  year={2002}
}

@article{Pena2006,
    title = {Nonstationary dynamic factor analysis},
    journal = {Journal of Statistical Planning and Inference},
    volume = {136},
    number = {4},
    pages = {1237--1257},
    year = {2006},
    doi = {https://doi.org/10.1016/j.jspi.2004.08.020},
    author = {Daniel Peña and Pilar Poncela}
}

@article{Bai2012,
    author = {Jushan Bai and Kunpeng Li},
    title = {Statistical analysis of factor models of high dimension},
    volume = {40},
    journal = {The Annals of Statistics},
    number = {1},
    pages = {436--465},
    year = {2012},
    doi = {10.1214/11-AOS966}
}

@article{Pena1987,
    author = {Daniel Peña and George E. P. Box},
    title = {Identifying a Simplifying Structure in Time Series},
    journal = {Journal of the American Statistical Association},
    volume = {82},
    number = {399},
    pages = {836--843},
    year = {1987},
    doi = {10.1080/01621459.1987.10478506}
}

@article{Blanchet2022,
    author = {Blanchet, Jose and Chen, Lin and Zhou, Xun Yu},
    title = {Distributionally Robust Mean-Variance Portfolio Selection with {W}asserstein Distances},
    journal = {Management Science},
    volume = {68},
    number = {9},
    pages = {6382--6410},
    year = {2022},
    doi = {10.1287/mnsc.2021.4155}
}

@book{galichon2016optimal,
  title={Optimal Transport Methods in Economics},
  author={Galichon, Alfred},
  year={2016},
  publisher={Princeton University Press}
}

@article{DeMarco2015,
    author = {De Marco, Stefano and Henry-Labord\`{e}re, Pierre},
    title = {Linking Vanillas and {VIX} Options: A Constrained Martingale Optimal Transport Problem},
    journal = {SIAM Journal on Financial Mathematics},
    volume = {6},
    number = {1},
    pages = {1171--1194},
    year = {2015},
    doi = {10.1137/140960724}
}

@article{Pesenti2023,
    author = {Pesenti, Silvana M. and Jaimungal, Sebastian},
    title = {Portfolio Optimization within a {W}asserstein Ball},
    journal = {SIAM Journal on Financial Mathematics},
    volume = {14},
    number = {4},
    pages = {1175--1214},
    year = {2023},
    doi = {10.1137/22M1496803}
}

@article{Santoro2023,
  title={Statistical inference for {B}ures-{W}asserstein flows},
  author={Santoro, Leonardo V and Panaretos, Victor M},
  journal={arXiv preprint arXiv:2310.13764},
  year={2023}
}

@book{Henry-Labordere2017,
    author = {Henry-Labordere, P.},
    title = {Model-free Hedging: A Martingale Optimal Transport Viewpoint},
    publisher = {Chapman and Hall/CRC},
    year = {2017} 
}

@book{dryden2016,
  title={Statistical Shape Analysis: With Applications in R},
  author={Dryden, I.L. and Mardia, K.V.},
  series={Wiley Series in Probability and Statistics},
  year={2016},
  publisher={Wiley}
}

@article{WANG2007,
    title = {A hyperspherical transformation forecasting model for compositional data},
    journal = {European Journal of Operational Research},
    volume = {179},
    number = {2},
    pages = {459--468},
    year = {2007},
    doi = {https://doi.org/10.1016/j.ejor.2006.03.039},
    author = {Huiwen Wang and Qiang Liu and Henry M.K. Mok and Linghui Fu and Wai Man Tse}
}

@article{frechet1948,
  author  = {Fr\'{e}chet, Maurice},
  title   = {Les \'{e}l\'{e}ments al\'{e}atoires de nature quelconque dans un espace distanci\'{e}},
  journal = {Annales de l'institut Henri Poincar\'{e}},
  volume  = {10},
  number  = {4},
  pages   = {215--310},
  year    = {1948}
}

@article{Zhang2022,
    author = {Zhang, Chao and Kokoszka, Piotr and Petersen, Alexander},
    title = {Wasserstein autoregressive models for density time series},
    journal = {Journal of Time Series Analysis},
    volume = {43},
    number = {1},
    pages = {30--52},
    doi = {https://doi.org/10.1111/jtsa.12590},
    year = {2022}
}

@article{Yu1994,
    author = {Bin Yu},
    title = {Rates of Convergence for Empirical Processes of Stationary Mixing Sequences},
    volume = {22},
    journal = {The Annals of Probability},
    number = {1},
    pages = {94--116},
    year = {1994},
    doi = {10.1214/aop/1176988849}
}

@article{Chen2023,
    author = {Yaqing Chen and Zhenhua Lin and Hans-Georg Müller},
    title = {Wasserstein Regression},
    journal = {Journal of the American Statistical Association},
    volume = {118},
    number = {542},
    pages = {869--882},
    year = {2023},
    doi = {10.1080/01621459.2021.1956937}
}

@article{Dai2022,
    author = {Xiongtao Dai},
    title = {{Statistical inference on the Hilbert sphere with application to random densities}},
    volume = {16},
    journal = {Electronic Journal of Statistics},
    number = {1},
    pages = {700--736},
    year = {2022},
    doi = {10.1214/21-EJS1942}
}

@article{Wilfrid2011,
    author = {Wilfrid S. Kendall and Huiling Le},
    title = {{Limit theorems for empirical Fréchet means of independent and non-identically distributed manifold-valued random variables}},
    volume = {25},
    journal = {Brazilian Journal of Probability and Statistics},
    number = {3},
    pages = {323--352},
    year = {2011},
    doi = {10.1214/11-BJPS141}
}

@article{Huckemann2006, 
    title={Principal component analysis for {R}iemannian manifolds, with an application to triangular shape spaces}, 
    volume={38}, 
    doi={10.1239/aap/1151337073}, 
    number={2}, 
    journal={Advances in Applied Probability}, 
    author={Huckemann, Stephan and Ziezold, Herbert}, 
    year={2006}, 
    pages={299--319}
}

@article{Huckemann2021,
    author = {Huckemann, Stephan F. and Eltzner, Benjamin},
    title = {Data analysis on nonstandard spaces},
    journal = {WIREs Computational Statistics},
    volume = {13},
    number = {3},
    pages = {e1526},
    doi = {https://doi.org/10.1002/wics.1526},
    year = {2021}
}

@article{Xu2025,
  author  = {Haoshu Xu and Hongzhe Li},
  title   = {Wasserstein {F}-tests for {F}r\'{e}chet regression on {B}ures-{W}asserstein manifolds},
  journal = {Journal of Machine Learning Research},
  year    = {2025},
  volume  = {26},
  number  = {77},
  pages   = {1--123}
}

@article{Xu2025test,
  title={Test of partial effects for {F}r\'{e}chet regression on {B}ures-{W}asserstein manifolds},
  author={Xu, Haoshu and Li, Hongzhe},
  journal={arXiv preprint arXiv:2506.23487},
  year={2025}
}

@inproceedings{Zirpoli2025,
    author = {Zirpoli, Michael and Yi, Yuyan and Lin, Shu-Chin and Ge, Linqiang and Zheng, Jingyi},
    title = {Towards Classification of Covariance Matrices via {B}ures-{W}asserstein-Based Machine Learning},
    year = {2025},
    doi = {10.1145/3719384.3719386},
    booktitle = {Proceedings of the 2024 7th Artificial Intelligence and Cloud Computing Conference},
    pages = {10–18},
    series = {AICCC '24}
}

@inproceedings{Haviv2025,
    title={Wasserstein Flow Matching: Generative Modeling Over Families of Distributions},
    author={Doron Haviv and Aram-Alexandre Pooladian and Dana Pe'er and Brandon Amos},
    booktitle={Forty-second International Conference on Machine Learning},
    year={2025}
}

@article{Jiang2025,
  title={Bures-{W}asserstein Flow Matching for Graph Generation},
  author={Jiang, Keyue and Cui, Jiahao and Dong, Xiaowen and Toni, Laura},
  journal={arXiv preprint arXiv:2506.14020},
  year={2025}
}

@INPROCEEDINGS{Tang2024,
  author={Tang, Hanning and Shen, Xiaojing and Zhao, Hua and Wang, Zhiguo and Varshney, Pramod K.},
  booktitle={2024 27th International Conference on Information Fusion (FUSION)}, 
  title={Bures-{W}asserstein Barycentric Coordinates with Application to Diffusion Tensor Image Smoothing}, 
  year={2024},
  pages={1--8},
  doi={10.23919/FUSION59988.2024.10706482}
}

@Article{zhu:24,
  author    = {Zhu, Changbo and M{\"u}ller, Hans-Georg},
  journal   = {Journal of Econometrics},
  title     = {Spherical autoregressive models, with application to distributional and compositional time series},
  year      = {2024},
  number    = {2},
  pages     = {105389},
  volume    = {239},
  doi       = {10.1016/j.jeconom.2022.12.008},
  publisher = {Elsevier},
}

@article{Markowitz1952,
    author = {Markowitz, Harry},
    title = {Portfolio selection},
    journal = {The Journal of Finance},
    volume = {7},
    number = {1},
    pages = {77--91},
    doi = {https://doi.org/10.1111/j.1540-6261.1952.tb01525.x},
    year = {1952}
}

@article{YEN2014,
    title = {Solving norm constrained portfolio optimization via coordinate-wise descent algorithms},
    journal = {Computational Statistics \& Data Analysis},
    volume = {76},
    pages = {737--759},
    year = {2014},
    doi = {https://doi.org/10.1016/j.csda.2013.07.010},
    author = {Yu-Min Yen and Tso-Jung Yen}
}

@book{tsay2010,
  author    = {Ruey S. Tsay},
  title     = {Analysis of Financial Time Series},
  edition   = {3},
  publisher = {John Wiley \& Sons},
  address   = {Hoboken, NJ},
  year      = {2010},
  doi       = {10.1002/9780470644560}
}

@article{Jung2012,
 author = {Sungkyu Jung and Ian L. Dryden and J. S. Marron},
 journal = {Biometrika},
 number = {3},
 pages = {551--568},
 title = {Analysis of principal nested spheres},
 volume = {99},
 year = {2012}
}

\end{document}